\documentclass[12pt]{article}
\usepackage{graphicx}
\usepackage{cite}
\usepackage{url}
\usepackage[utf8]{inputenc}
\usepackage{hyperref}
\usepackage{fancyvrb}
\usepackage{etoolbox}
\usepackage{amsmath}
\usepackage{amssymb}

\usepackage[compat=1.0.0]{tikz-feynman}
\usepackage{graphicx}
\usepackage{graphics}
\usepackage{caption}
\usepackage{tikz}
\usepackage{ctable}
\usepackage{dcolumn}
\usepackage{bm}
\usepackage{ifthen}
\usepackage{xspace}
\usepackage{rotating}
\usepackage{lineno,xcolor}
\usepackage{amsfonts}
\usepackage{booktabs}
\usepackage{capt-of}
\usepackage{color}

\usepackage{feynmp-auto}

\newcommand\pubnumber{}
\newcommand\pubdate{\today}

\def\kansas{Department of Physics and Astronomy\\
University of Kansas, Lawrence, KS 66045, U.S.A.}
\def\support{\footnote{Work supported by the National Science Foundation  
                       under awards PHY-1607262 and PHY-1913886.}}

\def\Title#1{\begin{center} {\Large #1 } \end{center}}
\def\Author#1{\begin{center}{ \sc #1} \end{center}}
\def\Address#1{\begin{center}{ \it #1} \end{center}}

\newcommand\pubblock{\rightline{\begin{tabular}{l} \pubnumber\\
         \pubdate  \end{tabular}}}
\newenvironment{Abstract}{\begin{quotation}  }{\end{quotation}}
\newenvironment{Presented}{\begin{quotation} \begin{center} 
             Talk presented on behalf of the ILD concept group at \end{center}\bigskip 
      \begin{center}\begin{large}}{\end{large}\end{center} \end{quotation}}
\def\Acknowledgments{\bigskip  \bigskip \begin{center} \begin{large}
             \bf Acknowledgments \end{large}\end{center}}

\def\beq{\begin{equation}}
\def\eeq#1{\label{#1}\end{equation}}
\def\eeqn{\end{equation}}

\def\beqa{\begin{eqnarray}}
\def\eeqa#1{\label{#1}\end{eqnarray}}
\def\eeqan{\end{eqnarray}}

\let\bar=\overbar

\def\Dslash{\not{\hbox{\kern-4pt $D$}}}
\def\dslash{\not{\hbox{\kern-2pt $\del$}}}

\def\msb{{\bar{\ssstyle M \kern -1pt S}}}

\RecustomVerbatimCommand{\VerbatimInput}{VerbatimInput}%
{fontsize=\footnotesize,
 frame=lines,  
 framesep=2em, 
 rulecolor=\color{gray},
 label=\fbox{\color{black}},
 labelposition=topline,
 commandchars=\|\(\) 
}

\begin{document}
\begin{titlepage}
\pubblock

\vfill
\Title{Study of WW $\rightarrow q\bar{q}\ell\nu$ at ILC500 with ILD }
\vfill
\Author{JUSTIN ANGUIANO\support}
\Address{\kansas}
\vfill
\begin{Abstract}


 This study showcases the approaches towards lepton identification and $\gamma\gamma$ mitigation at center-of-mass energy $\sqrt{s} = 500$ GeV for semileptonic WW decays at the ILC. The analysis is performed using fully simulated Standard Model Monte Carlo events with the ILD detector concept and emphasizes the measurement of the W mass. The mass measurement is performed through the identification of a lepton and treatment of the remaining system as the hadronic W-boson. Only the most favorable beam polarization scenario for WW production is considered. The resulting detector performance benchmark obtained, with an integrated luminosity of $1600 \, \, \text{fb}^{-1}$, is a statistical error on the W mass of $\Delta M_W\text{(stat.)} =  2.4 $ MeV and a relative statistical error on the WW cross-section of $\Delta \sigma / \sigma \text{(stat.)} = 0.036\% $.

\end{Abstract}
\vfill
\begin{Presented}
The International Workshop on Future Linear Colliders (LCWS 2019),\\
 Sendai, Japan, 28 October-1 November, 2019. C19-10-28.\\
\end{Presented}
\vfill
\end{titlepage}
\def\thefootnote{\fnsymbol{footnote}}
\setcounter{footnote}{0}

\AtEndEnvironment{thebibliography}{
}

\section{Introduction and Motivation}
\label{sec:Introduction}
The W-boson pair decaying semileptonically is a rich physics channel with a wide variety of facets to study. The W mass measurement is well motivated because it is an essential fundamental parameter of the standard model (SM). The WW channel is a unique way to pin down the W mass because the identification of the charged lepton automatically tags the hadronic W as the remaining particles in the system. The mass measurement quality is then only bounded by the performance of the detector and statistics. WW production is also sensitive to the polarization of electron positron collider beams, thus,  measuring the production cross-section provides an opportunity to implicitly measure the beam polarization at the interaction point. Another important aspect of WW, is the charged triple gauge couplings (TGCs). Deviation of these couplings from the standard model is a distinct signature of new physics. This study assesses the challenges associated with reconstruction and analysis of semileptonic W pairs at a center of mass energy of 500 GeV, which is an important tool in understanding the problems that lie in analysis within the next frontier of particle physics. The work presented contains four major steps. The first step is the identification of the lepton, which is done with a universal treatment of leptons, that is, without distinguishing between lepton flavors.  Secondly, the effects of $\gamma\gamma$ overlay on the hadronic mass are explored and a technique to reduce this effect is showcased. Next is the event selection where W pairs are selected against the full standard model background. Lastly, estimates for the statistical error of the W mass and cross-section are extracted.
\section{The W-boson}
\label{sec:physics}

\begin{figure}
\centering
\includegraphics[scale=0.24]{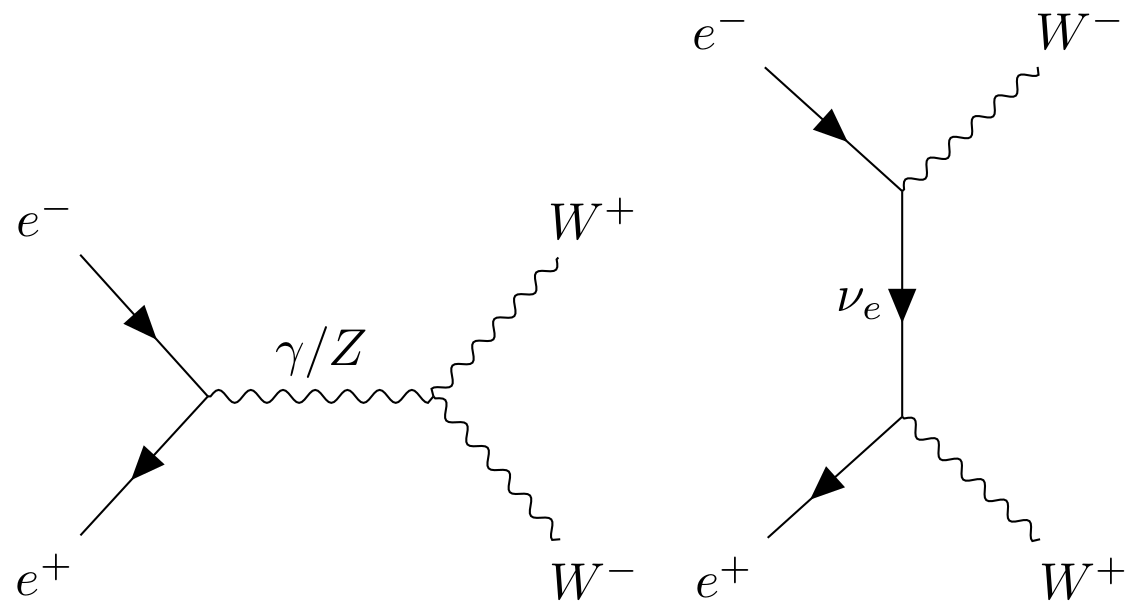}
\caption{\label{fig:wwdiag} Main WW production modes in the $s$ and $t$ channels. }
\end{figure}

 The current highest precision measurement for the mass and width are results of measurements from both WW production at $e^+ e^-$ colliders and single $W^\pm$ production at hadron colliders. These measurements use the combined results from LEP, Tevatron, and LHC experiments which report $M_W = 80.379 \pm 0.012 \, \, \text{GeV} $ and $\Gamma_W = 2.085 \pm 0.042 \,  \,\text{GeV}$ \cite{pdg}. The diagrams representing WW production are given in Figure \ref{fig:wwdiag}. The final states of the WW process are either the fully hadronic $WW\rightarrow q\bar{q}q\bar{q}$ which comprises $46\%$ of the total WW cross-section, semileptonic $WW\rightarrow q\bar{q}\ell\nu_{\ell}$ which comprises $44\%$, or fully leptonic $WW\rightarrow \ell \nu \ell \nu$ filling the remaining cross-section \cite{wwOPAL}. The semileptonic mode is the most favorable way to measure the W mass because the hadronic system is easily obtained after the identification of the lepton. The semileptonic channel also offers a large contribution to the $q\bar{q}e\nu_e$ final state through hadronic single W's alongside a non-resonant $e \nu_e$ pair. The single W contribution is overall less ``signal-like" to other $qq\ell\nu$ final states but assists in reducing the overall statistical uncertainty in a W mass measurement. The hadronic mode is more challenging due to the combinatoric assignment of the four hadronic jets into two W's along with color-reconnection which may cause ``cross-talk" between jets. The leptonic channel is also difficult because of the presence of two neutrinos and smaller branching fractions.

  The most difficult reconstruction of a final state from a semileptonic W decay is the case involving the tau lepton. The tau final states are equally as important as the light leptons because the W couples to the three lepton flavors democratically. The tau can mimic the signature of hadrons or other leptons in a detector in addition to producing additional missing energy via neutrinos.  The tau lepton mainly decays hadronically -- into a tau neutrino and virtual W-boson that produces a pair of quarks. The virtual W's daughter quarks will hadronize into a charged particle ($\pi^\pm$) or form more charged or neutral particles ($\pi^0$). The virtual W in the tau decay is allowed to couple with leptons, so, the tau final state can include either an electron or muon along with the corresponding flavor neutrinos. The decay rates for the tau are given in Table \ref{tab:taudecay}.
\begin{table}
\centering
 \begin{tabular}{|c|l|c|} 
 
 \hline
       & Decay Mode & Branching Ratio  \\ \hline \hline
    Hadronic Modes  & $\pi^- \nu_\tau$  & $10.82\%$  \\
      	($64.79\%$) & $\pi^- \pi^0 \nu_\tau$ & $25.49\%$ \\
     				& $\pi^- \pi^0 \pi^0 \nu_\tau$  & $9.26\%$  \\
     				& $\pi^- \pi^0 \pi^0 \pi^0 \nu_\tau$  & $1.04\%$   \\
      				& $\pi^- \pi^+ \pi^- \nu_\tau$  & $8.99\%$      \\ 
      				& $\pi^- \pi^+ \pi^- \pi^0 \nu_\tau$  & $2.74\%$  \\ \hline
    			    
    Leptonic Modes  & $e^- \nu_e \nu_\tau$ & $17.82\%$   \\
    	($35.21\%$)	& $\mu^- \nu_\mu \nu_\tau $  & $17.39\%$      \\ \midrule \hline

\end{tabular}
        \caption{\label{tab:taudecay}Most common decay modes for the $\tau^-$ lepton \cite{pdg}}
\end{table}

  Measurements in the WW channel through  $e^+ \, e^-$ annihilation presents two important effects: (1) contributions from foreign particles which are reconstructed alongside the primary interaction and (2) the rate at which the W-pairs are produced, which is sensitive to beam helicities. The first effect has a contribution from two components, beamstrahlung and overlay. In beamstrahung photons are produced from the interactions between the fields of the beams. The radiated photons generally go undetected by escaping down the beam-pipe causing the effective center of mass energy to be reduced at the interaction point. The photons interact with other photons at a rate of 1.1 events per beam crossing \cite{ILDIDR} creating what is known as the $\gamma \gamma$ overlay. The $\gamma \gamma \rightarrow \text{hadrons}$ may scatter into the detector and add a source of confusion wherein the foreign particles ``overlay" on top of the true particles of an event. For the second effect, the W-pair production rate, there are four possible combinations of electron and positron helicities where each initial particle is either left or right handed. More explicitly, a collision can consist of  $e^-_L e^+_R$(LR) with left-handed electron and right-handed positron , $e^-_R e^+_L$(RL)  with right handed electron and left handed positron, or mirroring helicities RR and LL.  The beams in the collider are mixed with multiple helicities which is represented by overall partial longitudinal beam polarizations $P_{e^-}$ and $ P_{e^+}$.  Two W-bosons can only be produced in LR and RL configurations, whereas in the mirrored configurations, the recombination into a particle of spin 1 is  not possible.  The W has pure coupling only to left handed electrons or right handed positrons in the $t$-channel, so, the number of WW events produced are sensitive to the beam polarization\cite{thomson}.  If the number of events produced is sensitive to polarization then the overall cross-section for WW production is sensitive to beam polarization.



\section{Event Generation and Detector Simulation}
\label{sec:ILC_detector}

The software ecosystem for the ILC is contained under iLCSoft \cite{ilcsoft} version \url{v02-00-02} which is comprised of reconstruction tools that rely on the event data model LCIO\cite{lcio}. Full simulation samples that are generated are based on detector descriptions in DD4HEP \cite{dd4hep}. The physics events are centrally produced with the Whizard event generator \cite{ whizard} with hadronization performed in Pythia6 \cite{pythia6} then simulated in the detector with Geant4 \cite{geant4}. The Monte Carlo samples being used are for the Interim Design Report benchmarking effort \cite{ILDIDR} and are based on generator level events created for the ILD Detailed Baseline Design \cite{tdrdet}. The analysis relies on Monte Carlo events that are fully simulated using the ILD detector model \url{ ILD_l5_o1_v02 } and includes a complete standard model background for final states with 2, 4, and 6 fermions as well as standard model Higgs production.  The Monte Carlo events are generated at a center of mass energy of 500 GeV with $100\%$  longitudinally-polarized beams. The $\gamma \gamma$ event contributions are overlaid on top of the hard event before reconstruction. Events are reconstructed with PandoraPFA \cite{particleflowA} into particle flow objects (PFOs). Each event is weighted in order to obtain a realistic case of partial polarization for a possible running with $-80\%$ for the electron beam and $+30\%$ for the positron beam, comprising an integrated luminosity of 1600 $\text{fb}^{-1}$.  The analysis framework for this project can be found at \cite{wwrepo}.


\section{Measurement of the W mass and Cross-section}
\label{Current_Work} 

\subsection{Analysis Overview}
\label{subsec:ana_overview}

The analysis workflow for the semileptonic mode has three distinct stages, the lepton identification and selection, overlay rejection in the hadronic system, and event selection against full standard model backgrounds. The analysis is performed with the  running scenario that creates the most dominant WW production mode $(-0.8,+0.3)$.

\subsection{Lepton Identification}
\label{subsec:Lepton_ID}
The approach towards the identification of leptons relies on treating leptons universally. The easiest lepton to identify is the muon, which produces a single track along with hits in the muon detector. The electron also produces a track in the TPC but is often accompanied by photons via bremsstrahlung. The tau is the most difficult lepton to identify due to its frequent decay into multiple charged and neutral particles. To accommodate all types of lepton signature, a cone based approach is used to either capture single tracks or collimated jets with low track multiplicity. The lepton finding cone is based on the TauFinder package \cite{taufinder} developed initially for the Compact Linear Collider (CLIC) studies. TauFinder consists of two major structures, a search cone containing the particles that belong to the lepton candidate and an isolation cone whose purpose is to reject a lepton candidate if the search cone is not well isolated from other particles. The acceptance criteria for a search cone consists of these parameters:
\begin{itemize}
\item Search cone angle $\alpha$ - The opening (half) angle of the search cone for the lepton jet [rad]
\item Isolation cone angle $\beta$ - The outer isolation cone angle with respect to the search cone [rad]
\item Isolation energy - The total energy allowed within the isolation cone region [GeV]
\item Invariant mass - The upper limit on the lepton candidate mass [GeV]
\item Track multiplicity - The allowed number of tracks in a lepton candidate
\item Minimum $P_T$ seed - the minimum transverse momentum of a track that seeds a lepton candidate [GeV] 
\end{itemize}
An example of the cone and parameters are shown in Figure 2. Additional requirements are imposed on all of the reconstructed Particle Flow Objects (PFOs) in the event in order to suppress overlay particles being included in the lepton jet.
\begin{itemize}
\item $P_T > 0.2$ GeV
\item $|\cos\theta| < 0.99$
\end{itemize}
The formation of a lepton candidate follows three steps (1) candidate construction, (2) candidate merging, and (3) isolation testing.
The first step starts with seed tracks that are sorted by energy in descending order. Any track or neutral particle that falls within the search cone of the lepton candidate is added to the lepton candidate. For each newly added particle, the energy and momentum of the lepton candidate is updated. Each candidate has a unique set of particles. Lepton candidates are continually formed until the seed tracks are exhausted. When there are no more candidates to be created, the candidates are subjected to part of the acceptance criteria: the lepton jet mass is required to be below the upper mass limit (2 GeV) and the number of charged tracks within the lepton candidate is non-zero and less than or equal to 4. If a lepton jet violates any acceptance condition it is deleted. The next step in the process is merging. If two lepton candidates form an opening angle of less than $2\alpha$, the candidates are merged. If the mass or track multiplicity conditions are violated, both lepton candidates are deleted.  All  candidates that survive merging are subjected to the isolation testing. For each candidate, the sum of energy of all the particles that fall inside the isolation cone is computed. If the total energy inside the isolation cone is greater than the maximum allowed energy inside the isolation cone the lepton candidate is deleted.\\

	The universal lepton treatment is not conducive to a one-size-fits-all approach to lepton ID due to the abundance of different lepton signatures. To accommodate variations between lepton signatures, the acceptance criteria for leptons is optimized according to lepton flavor and $\tau$ decay mode. The categories created are: Prompt $\ell$, $\tau \rightarrow \ell\bar{\nu_{\ell}} \nu_{\tau}$, $\tau \rightarrow $hadrons. The leptonic categories, for both prompt and $\tau$ decays, are further subdivided between light flavors whereas the hadronic category is separated by either 1-prong or 3-prong decays.\\

\begin{figure}[hp!]
\centering
\captionsetup{justification=raggedright,margin=3cm}
\includegraphics[width=0.4\textwidth]{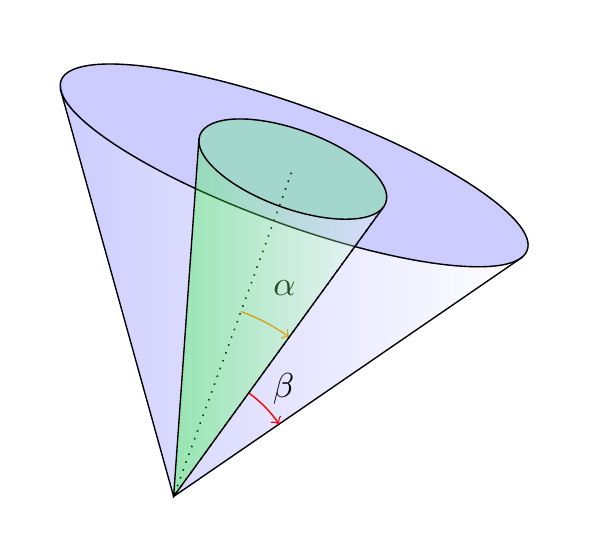}
\caption{Illustration of possible lepton candidate cone with search cone angle $\alpha$ and isolation cone angle $\beta$. The search cone is shown in green and the isolation cone is the surrounding blue cone.}
\end{figure}

\begin{figure}
\centering
    \begin{minipage}{0.48\textwidth}
        \centering

\begin{tikzpicture}
    \node(a){\includegraphics[width=0.99\textwidth]{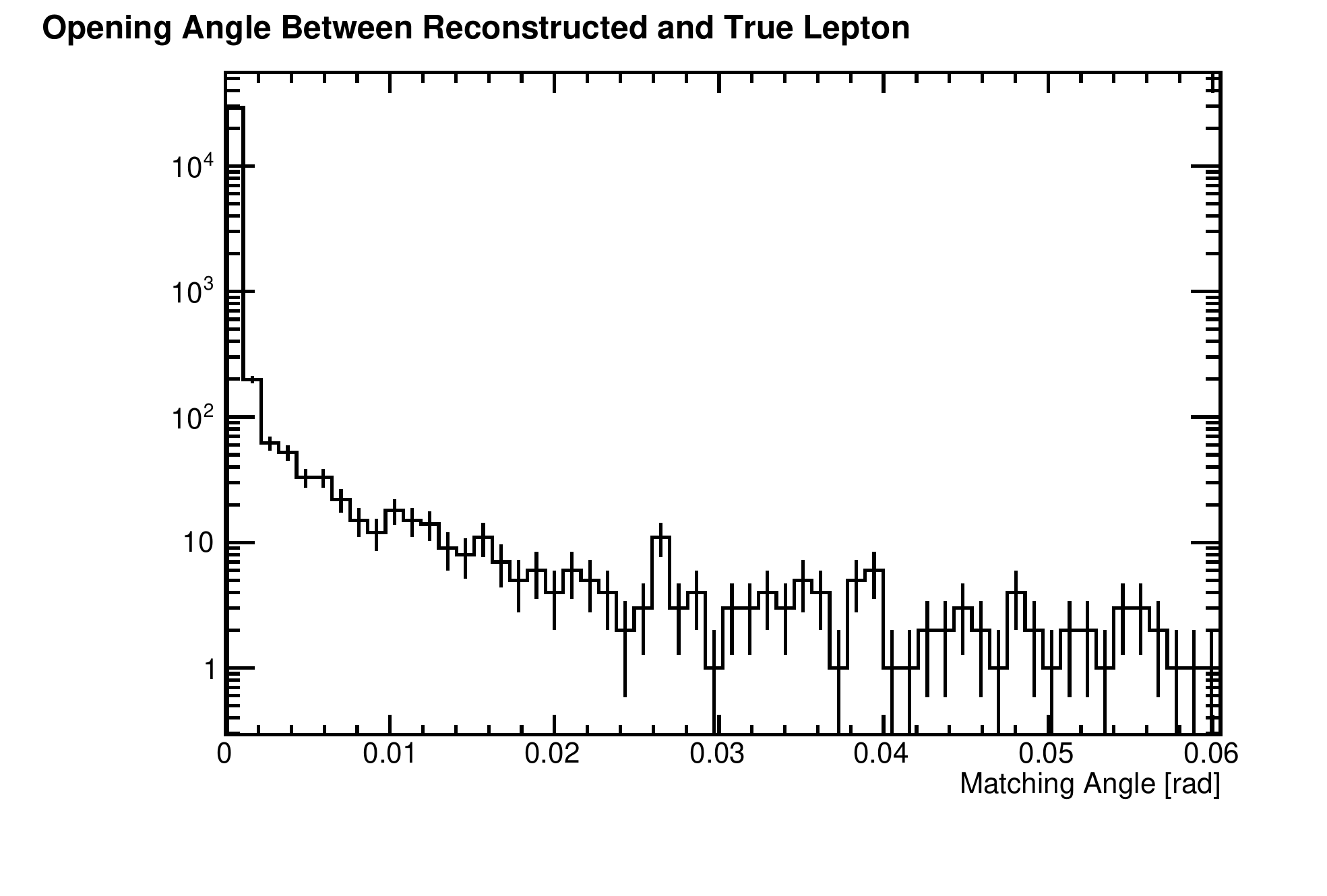}};
    \node at (a.north west)
    [
    anchor=center,
    xshift= 2.5cm,
    yshift= -1cm
    ]
    {
        \includegraphics[width=0.13\textwidth]{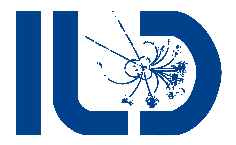}
    };
    \end{tikzpicture}
\caption{Distribution of opening angles between the closest reconstructed lepton candidate and the true muon from $WW \rightarrow q \bar{q} \mu \nu_\mu$. $99.4\%$ of events with a muon candidate are matched to truth.} 
\label{fig:taupsi}
\end{minipage}\hfill
    \begin{minipage}{0.48\textwidth}
        \centering

\begin{tikzpicture}
    \node(a){\includegraphics[width=0.99\textwidth]{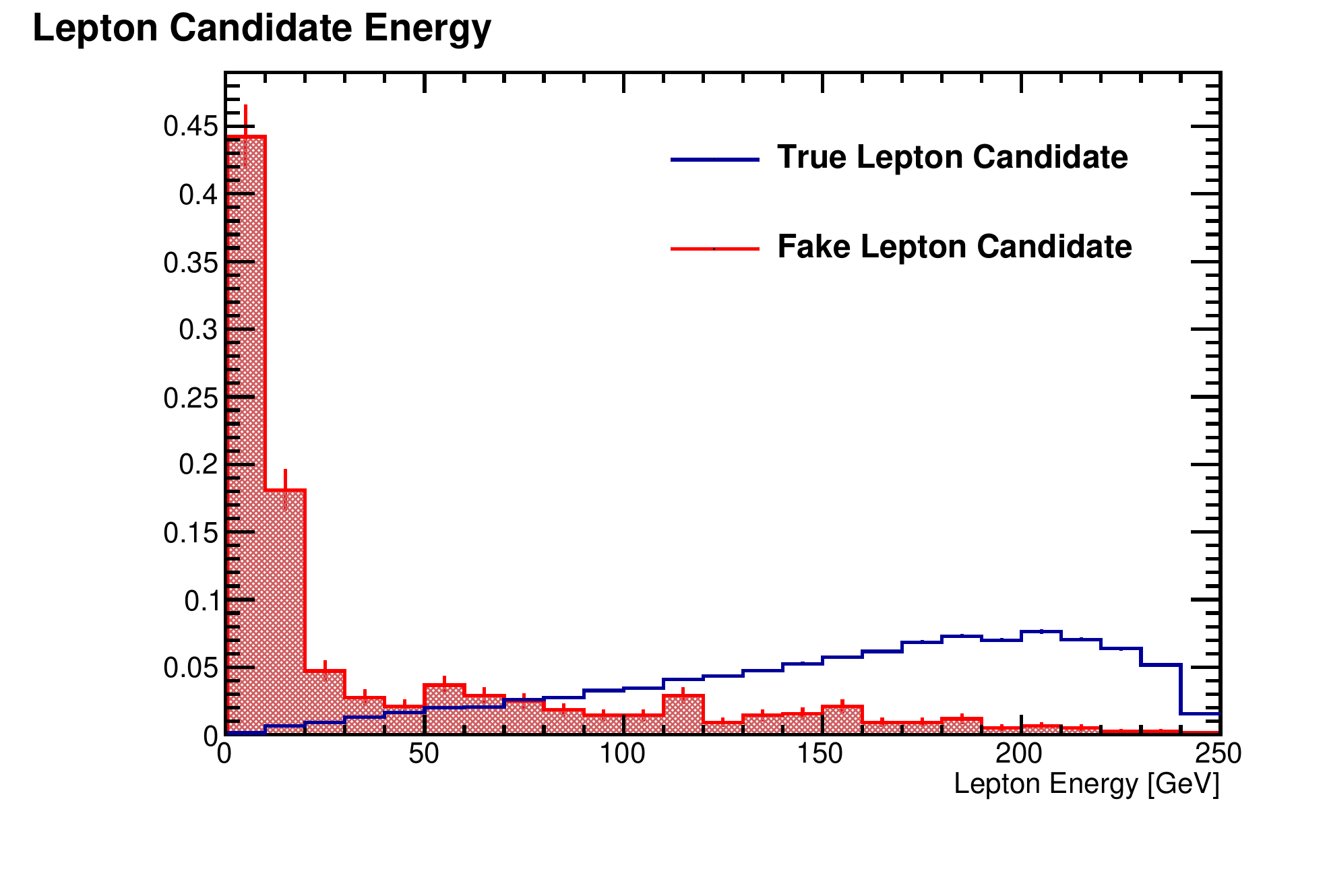}};
    \node at (a.north west)
    [
    anchor=center,
    xshift= 2.5 cm,
    yshift= -1cm
    ]
    {
        \includegraphics[width=0.13\textwidth]{ildlogo.png}
    };
    \end{tikzpicture}
\caption{Energy distribution of lepton candidates matched to truth from $WW \rightarrow q \bar{q} \mu \nu_\mu $ and fake candidates from $ WW \rightarrow q\bar{q} q \bar{q}$ both normalized to unity.\\}
\label{fig:candE}
\end{minipage}
\end{figure}

 The optimal selection criteria for each category is the set of parameters that maximally identify lepton candidates that originate from true leptons and minimize the fake lepton candidates that originate from hadronic jets. To find this set of parameters, a scan over a 3D space is performed using the search cone-$\alpha$, isolation cone-$\beta$, and isolation energy-$E_{\text{iso}}$. The invariant mass upper limit is held at a fixed 2 GeV for simplicity.
Two parameters are defined to find the optimal working point in the lepton finding space. The first is related to correctly identifying jets originating from true leptons.
The true lepton reconstruction efficiency is maximized with the signal sample $WW\rightarrow q\bar{q}\ell\nu$ and denoted as  $\epsilon_T$.
\begin{equation}
\label{eq:et}
\epsilon_T = N_{\text{match}}/N_{\text{Stotal}}
\end{equation}
 The denominator, $N_{\text{Stotal}}$, represents the total, category specific, number of events which contain three generator visible fermions. The truth $q\bar{q}\ell$ fermions are required to fall within the acceptance range $|\cos\theta| < 0.99$. $N_{\text{match}}$ is the number of signal sample events in which a lepton candidate is reconstructed and can be matched to the true lepton, such that the opening angle between the reconstructed lepton and the true lepton is less than 0.1 radians. The distribution of opening angles is shown in Figure \ref{fig:taupsi}. In the case that a reconstructed lepton is being matched to a true tau, the matching angle is formed between the reconstructed lepton and the vector sum of the visible generator components of the tau decay. The visible components of the tau decay consist of the direct decay products whereas photons from final state radiation of the tau prior to decay are excluded.

 The second optimization parameter is denoted as $P_F$, the probability of a fake lepton jet arising from a single hadronic jet. 
\begin{equation}
\label{eq:pf}
P_F = 1-(1-\epsilon_F)^{\frac{1}{4}} 
\end{equation} 
 The fake lepton probability $P_F$ is minimized using the background sample $WW\rightarrow q\bar{q}q\bar{q}$ and is a function of the fake lepton reconstruction efficiency $\epsilon_F$.
\begin{equation}
\label{eq:ef}
\epsilon_F = N_{\text{fake}}/N_{\text{Btotal}}
\end{equation}
The $\epsilon_F$ denominator, $N_{\text{Btotal}}$ represents the total number of background events and is also subjected to the same acceptance range $|\cos\theta| < 0.99$ for all four fermions. The numerator, $N_{\text{fake}}$, is the total number of events  that contain at least one reconstructed fake lepton. The fake efficiency can be interpreted as the binomial probability of $r$-successes (lepton reconstructions) in 4 trials (hadronic jets). The probability of a single success in a single trial, $P_F$, can be directly derived from the binomial p.d.f using the fake efficiency $\epsilon_F$ per hadronic jet. 
The optimal parameters $\alpha$, $\beta$, $E_{\text{iso}}$ for each lepton category are extracted from max$[\epsilon_T(1-P_F)]$. The results for each category are shown in Table \ref{tab:taufinderopt}. 

\begin{table}

\begin{tabular}{|p{0.14\textwidth}|p{0.14\textwidth}p{0.14\textwidth}p{0.14\textwidth}p{0.1\textwidth}p{0.1\textwidth}p{0.1\textwidth}|}

\hline 
Channel & $n\,\text{Lep}\geq 1$ $\,\, (\%)$  & $\epsilon_T \, \, (\%)$ & $1-P_{F} \, \, (\%)$ & $\alpha$ [rad] & $\beta$ [rad] & $E_{\text{iso}}$ [GeV] \\ 
\hline 
Prompt $\mu$ & $95.5 \pm 0.3$  & $94.9 \pm 0.3$ & $97.4 \pm 0.1$& 0.03 & 0.15 & 3.0 \\ 

Prompt $e$ & $92.0 \pm 0.3$  & $90.4 \pm 0.3$ & $96.1 \pm 0.1$ & 0.04 & 0.15 & 4.0 \\ 

Inclusive $\tau$ & $80.0 \pm 0.5$  &  $77.0 \pm 0.6$ & $94.3 \pm 0.1$& 0.07 & 0.15 & 4.5 \\

 \hline
$\tau \rightarrow \nu \nu \mu$ & $81.5 \pm 1.2$  & $80.1 \pm 1.3$ & $97.4 \pm 0.1$ & 0.03 & 0.15 & 3.0 \\ 
 
$\tau \rightarrow \nu \nu e$  &  $80.0 \pm 1.2$&   $78.1 \pm 1.3$ & $96.3 \pm 0.1$  & 0.05 & 0.15 & 3.5 \\ 
 
$\tau$ Had-1p & $74.4 \pm 0.9$  & $70.7 \pm 0.9$ & $93.0 \pm 0.2$ & 0.07 & 0.15 & 4.5 \\ 
 
$\tau$ Had-3p &  $75.6 \pm 1.5$  & $71.0 \pm 1.6$ & $93.0 \pm 0.2$ & 0.07 & 0.15 & 5.5  \\
\hline
\end{tabular} 
\caption{Optimization results using $100 \%$ LR $q\bar{q}\ell \nu$ and $q\bar{q} q\bar{q}$ samples. The $n \, \, \text{Lep} \geq 1$ column pertains to signal samples where at least one lepton candidate was found and is not subjected to the truth matching criterion of 0.1 radians. Results shown are the configurations that maximize $\epsilon_T(1-P_F)$. The Prompt $\mu$ and Inclusive $\tau$ cones are chosen for a tight and loose lepton selection respectively. }
\label{tab:taufinderopt}
\end{table}

Since only one lepton is expected from signal, a single lepton candidate is selected as the candidate for the event. If multiple lepton jets are reconstructed then the lepton candidate with the highest energy is selected as the single candidate for the event. Any additional lepton candidates are treated as part of the hadronic system. The energy distribution of true and fake leptons is shown in Figure \ref{fig:candE}.

\subsection{$\gamma\gamma$ Overlay Mitigation}
\label{subsec:Pileup_mitigation}
\begin{figure}

\includegraphics[width=0.48\textwidth]{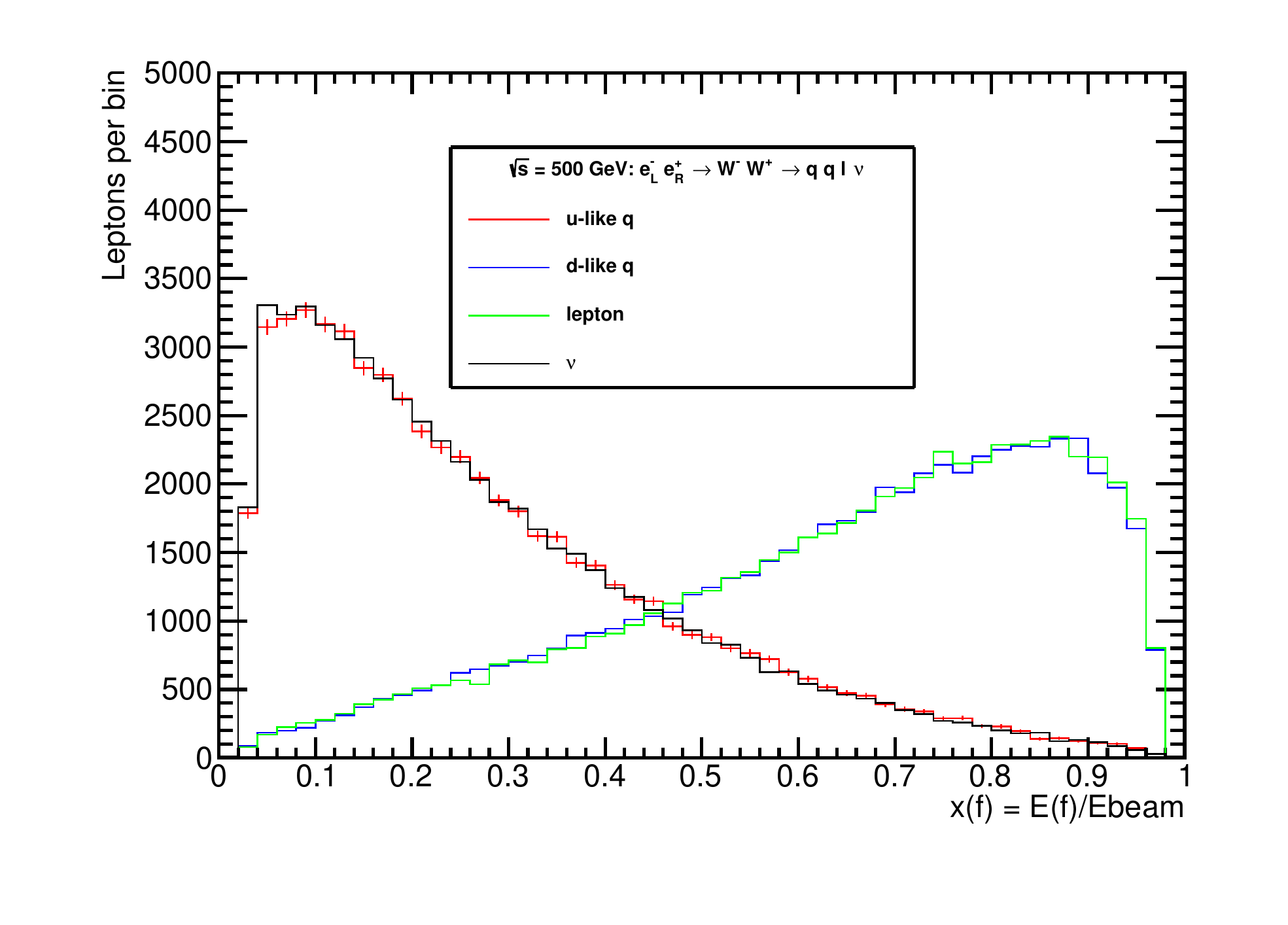}
\includegraphics[width=0.48\textwidth]{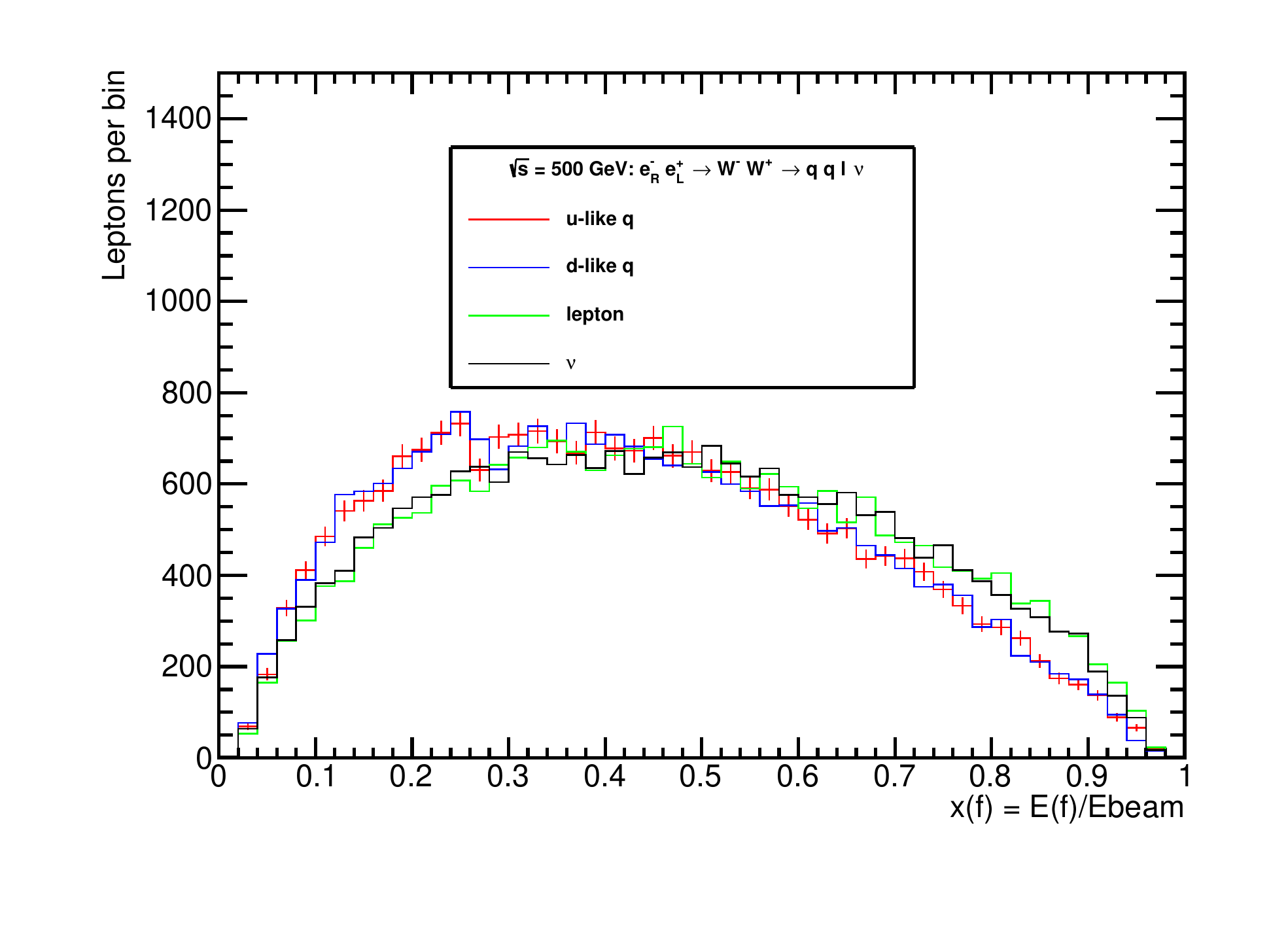}
\caption{The fractional energy partitioning of the true fermions with $\ell = \mu,\tau$, $100\%$ polarization for the initial state helicities LR (Left) and RL (Right) at center of mass energy 500 GeV. In the LR configuration the charged lepton and down-like quark take the majority of the beam energy. In RL configuration the energy partitioning is similar among the four fermions. }
\label{fig:Epartition}
\end{figure}

\begin{figure}

\includegraphics[width=0.48\textwidth]{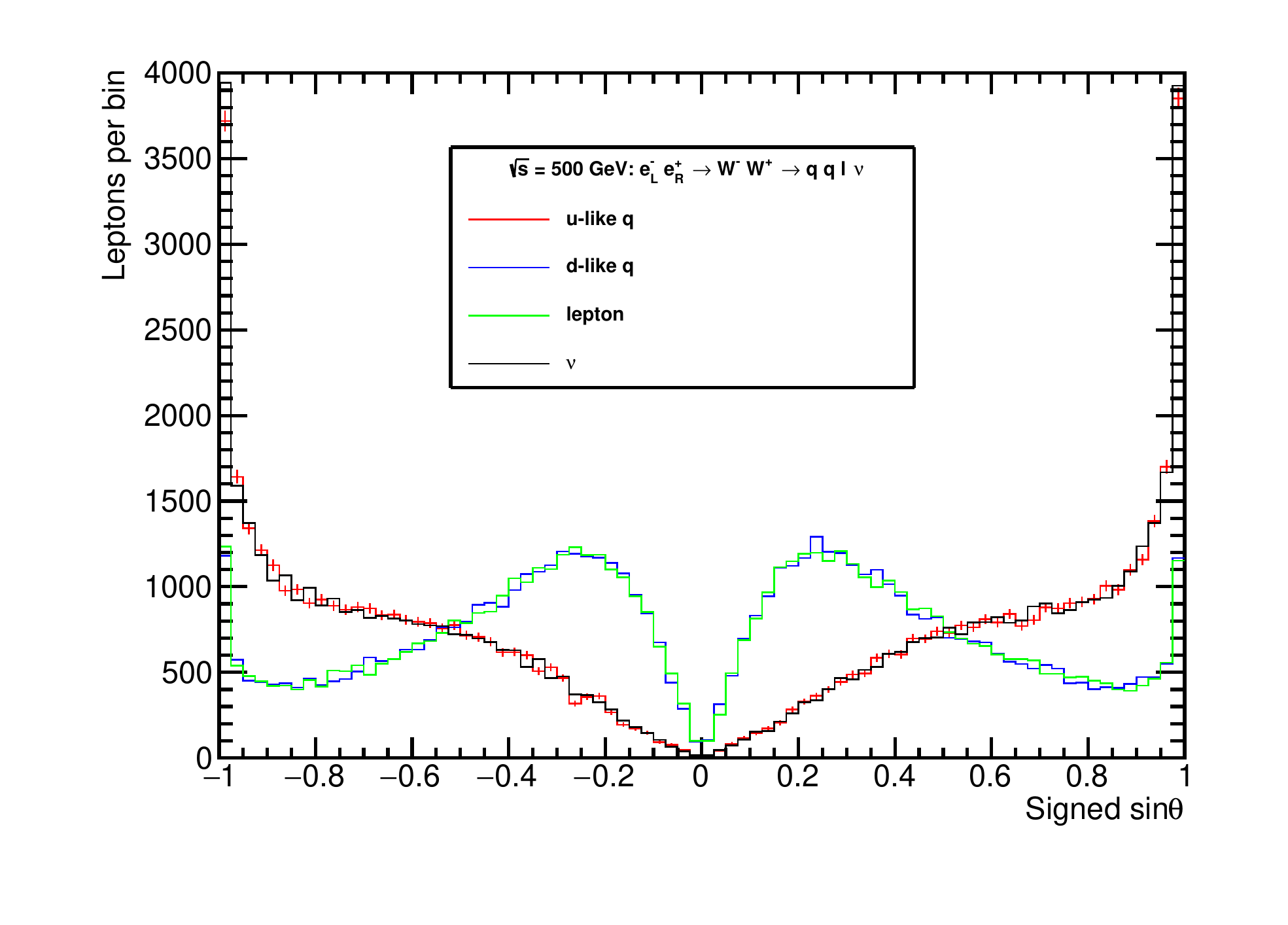}
\includegraphics[width=0.48\textwidth]{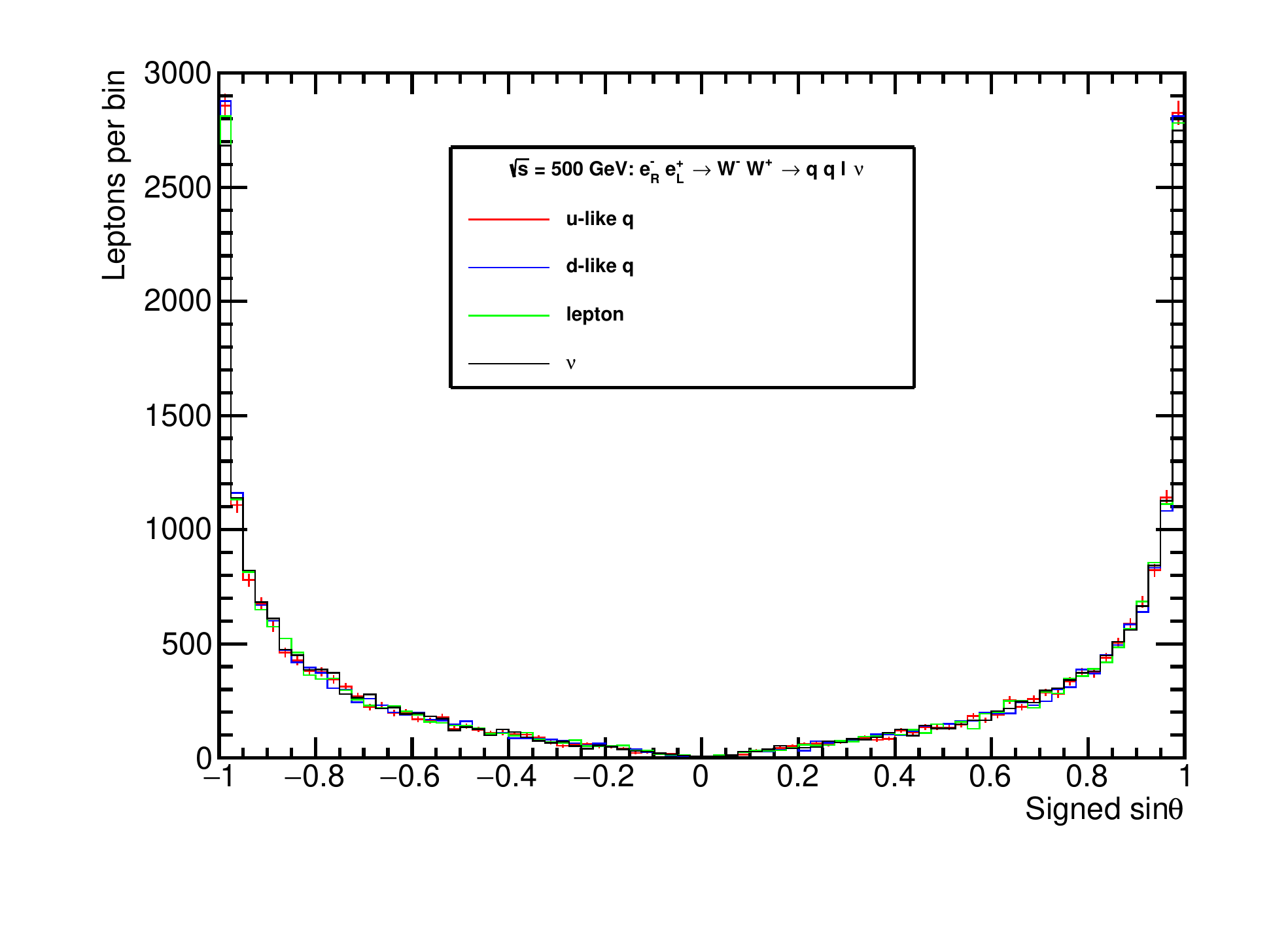}
\caption{The Signed sine of the polar angle of the true fermions with $\ell = \mu,\tau$. $100\%$ polarization for the initial state helicities. The sign of $\sin\theta$ corresponds to the sign of $\cos\theta$. As $\sin\theta \rightarrow 0$ the fermion is forward in the detector but for $|\sin\theta| \rightarrow 1$ the fermions become maximally transverse to the beam. In the LR configuration (left) the charged lepton and down-like quark are scattered forward while the up-like quark and neutrino are ejected centrally into the detector. In the RL configuration (right) all of the fermions are more centrally produced. }
\label{fig:fangles}
\end{figure}

\begin{figure}
    \centering
        \centering

        \begin{tikzpicture}
   \node(a){\includegraphics[scale=.45]{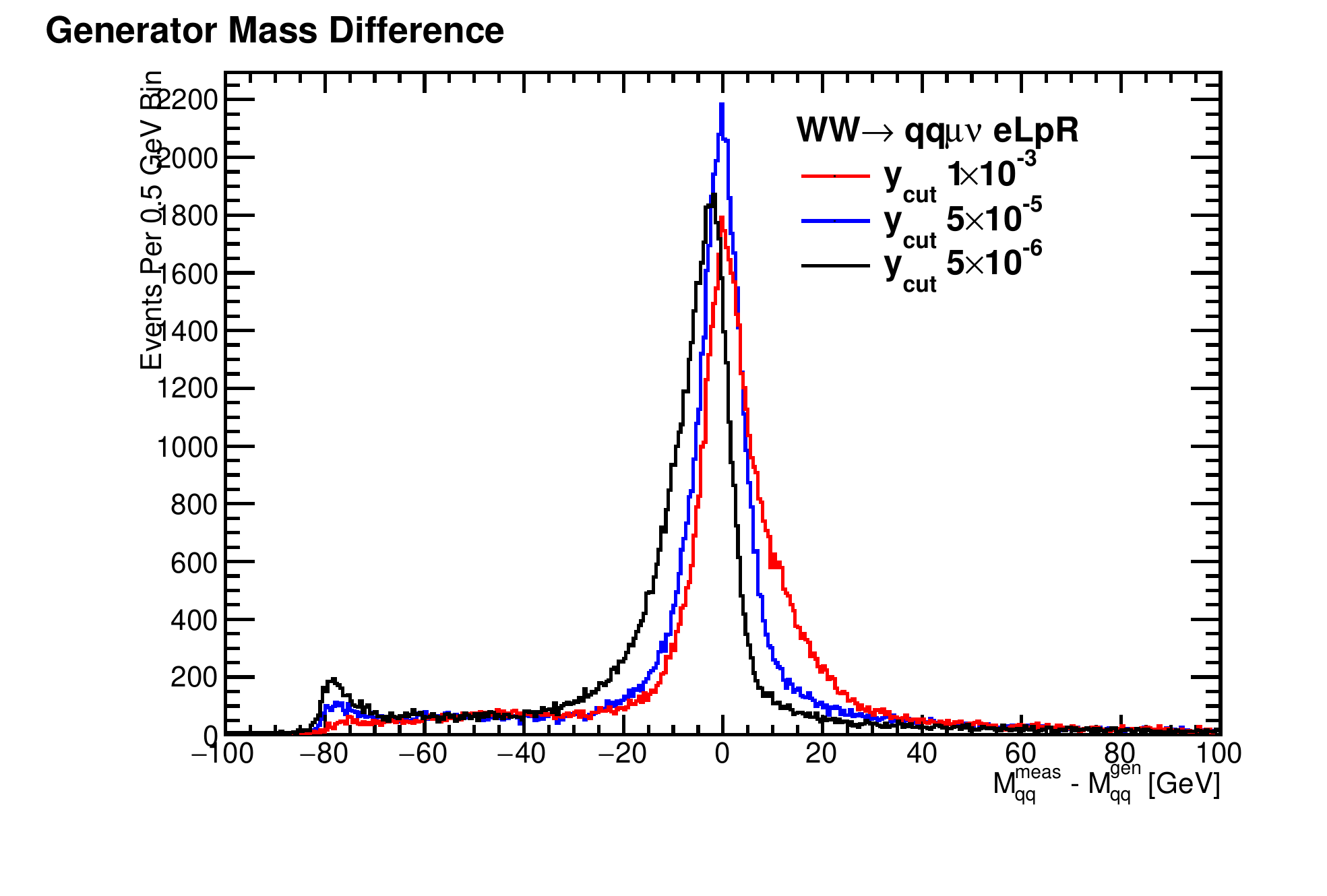}};
    \node at (a.north west)
    [
    anchor=center,
    xshift= 3.3cm,
    yshift= -1.3cm
    ]
    {
        \includegraphics[width=0.06\textwidth]{ildlogo.png}
    };
    \end{tikzpicture}
        \caption{Comparison of generator and reconstruction mass differences with different $y_{\text{cut}}$ values and the same  flat mini-jet cut of $P_T > 2$. The generator mass is computed from the vector sum of true di-quark pair from the hadronic W boson. The results of the largest $y_{\text{cut}}$ are massive (high $P_T$) jets that are not separated from overlay and insensitive to small kinematic cuts. $y_{\text{cut}} = 5\times 10^{-5}$ has the best balance between jet clustering and mini-jet cuts. $y_{\text{cut}} = 5\times 10^{-6}$ yields a highly fragmented system where the mini-jets are not distinguishable from overlay. The most fragmented system is sensitive to the $P_T$ requirement resulting in the small peak around -80 GeV where the hadronic W is completely thrown out.  }
        \label{fig:supdiff}
       \end{figure}

\begin{figure}
        \centering

        \begin{tikzpicture}
    \node(a){\includegraphics[scale=.45]{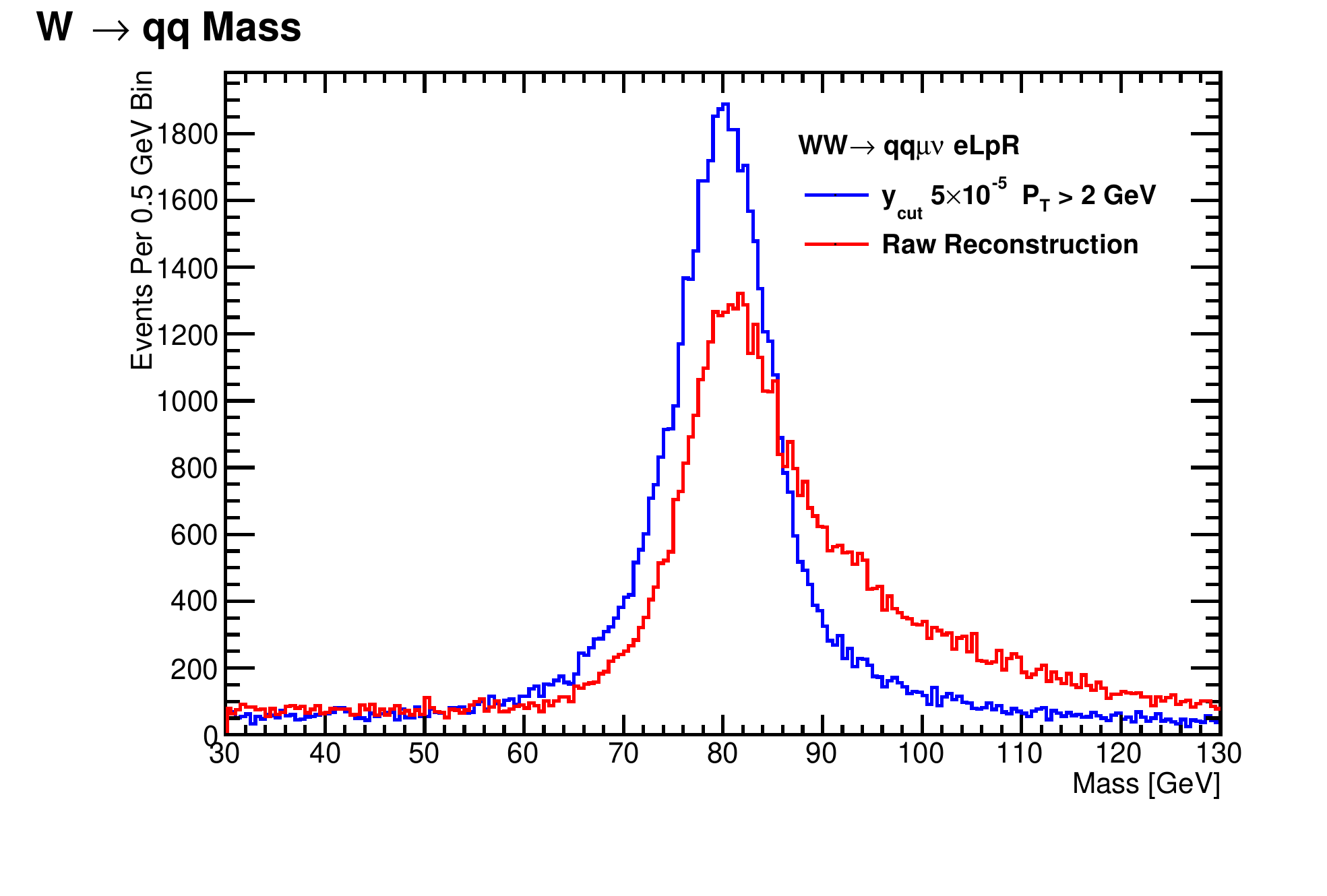}};
    \node at (a.north west)
    [
    anchor=center,
    xshift= 3.3cm,
    yshift= -1.3cm
    ]
    {
        \includegraphics[width=0.06\textwidth]{ildlogo.png}
    };
    \end{tikzpicture}
        \caption{The increase of quality of the hadronic mass is shown between the red curve which is the raw hadronic system after lepton identification versus the black curve which is subjected to the overlay mitigation with $y_{\text{cut}} = 5\times10^{-5}$ and mini jet $P_T > 2$ GeV. On average the excess in mass is reduced by $\approx 5$ GeV. }
        \label{fig:supmass}
\end{figure}

After a lepton candidate has been selected, the remaining particles in the system are expected to form the hadronically decaying W boson. However, hadronic mass is often in excess of the true hadronic mass. Variation between the true and measured mass naturally arises due to the mismeasurement of particles -- especially neutral hadrons, as well as particles lost beyond the acceptance of the detector. This effect is not substantial enough to account for the systematic excess observed in reconstruction. The nature of the excess in mass can be understood through the kinematics of the WW in the LR and RL configurations shown in Figures \ref{fig:Epartition} and \ref{fig:fangles}. The highest yielding configuration, LR, typically has two fermions that are forward in the detector which both typically have a large fraction of the beam energy. These fermions are susceptible to $\gamma\gamma$ scattering into the detector and mixing directly into the reconstructed jets. To combat effects of the $\gamma \gamma$ overlay, jet clustering algorithms via FastJet\cite{fastjet} are used.  The standard approach for overlay mitigation is to use the $k_T$ algorithm\cite{kt} and tune the R parameter such that the overlay particles are associated with beam jets while the desired particles are not. With successful $k_T$ clustering the beam jets can be thrown away without damaging the reconstruction of the desired event. However, this approach only works well in events that are centrally produced.  The overlay overlap in the forward topology with the $k_T$ algorithm  based clustering leads to rejecting desired particles and severe undermeasurment of the W mass. The solution to proper overlay mitigation is through the precise removal of foreign particles inside the reconstructed jets.  This can be achieved by using the standard JADE algorithm and cut-off parameter $y_{\text{cut}} > y_{ij}$ where $y_{ij} = d_{ij} / Q^2$ with $d_{ij} = 2E_i E_j(1-\cos\theta_{ij})$ \cite{fastjet}.  The mass of individually reconstructed jets can be controlled by tuning the $y_{\text{cut}}$ parameter. For large values, $y_{\text{cut}} =1\times10^{-3}$, a single massive jet is reconstructed. In the limit that $y_{\text{cut}}$ becomes infinitely small the number of jets reconstructed converges to the number of reconstructed particles.  The best $y_{\text{cut}}$ value is the value that forms mini-jets that safely couple together hard and soft emissions from  the original quark jet while segregating overlay into its own mini-jets. The mini-jets are then subjected to kinematic cuts that maximize the overlay rejection and minimize the difference  between the true and measured hadronic W mass. 	The best combination of $y_{\text{cut}}$ and mini-jet kinematic cuts are found by examining the $100\%$ polarized LR signal dataset for $qq \mu \nu$.  Two statistical estimators are used to maximize the overlay rejection, both of which come from the distribution of $M_{qq}^{\text{meas}} - M_{qq}^{\text{true}}$. This binned mass difference distribution is created from the subset of mini-jets that arise from clustering with a given $y_{\text{cut}}$ and also pass some jet requirements $P_{T} > x$ and $|\cos\theta| < y$. The estimators, from the distribution, are the Full Width Half Maximum (FWHM) and the number of entries in the Mode.  
Using estimators calculated from a binned histogram creates unwanted sensitivity to bin size. To reduce sensitivity to binning, firstly, the mode is defined as the bin with the most entries. The ``mode entries" is defined as the number of entries in the mode bin plus the number of entries in the nearest neighbors of the mode bin. For the FWHM, the mass distribution is assumed to be monotonically decreasing around the half maximum. To create a more sensitive continuous distribution of the FWHM,  the FWHM is weighted towards the bin center of the two bins around (above/below) the half maximum. The results of the optimization are shown in Figure \ref{fig:supmass} and various $y_{\text{cut}}$'s are shown in comparison to the optimal configuration in Figure \ref{fig:supdiff}. The optimal result uses $y_{\text{cut}} = 5\times 10^{-5}$, mini jet $P_T > 2$ GeV, and has no $|\cos \theta|$ requirement. 

\subsection{Event Selection}
\label{subsec:EventSelection}
The W-pair selection has been optimized for a Monte Carlo sample of 1600 $\text{fb}^{-1}$ with the $(-0.8,+0.3)$ beam scenario and builds on those described in \cite{ivan}. However, there is a significant difference between the present and preceding analysis selections, such that  the previous selection only addresses prompt muons as signal and treats muonic tau decays as background. The selection includes the full 2, 4, 6 fermion and Higgs SM background, and is performed with two mutually exclusive subsets, a tight and loose selection. The tight selection uses the prompt muon cone to identify signal events that contain both prompt and non-prompt muons and electrons. The tight selection is inefficient in collecting hadronic taus, so, the loose selection, using the inclusive tau cone, is designed to recover the efficiency of hadronic taus and other problematic events. The tight and loose cone parameters are given from the optimization in Table \ref{tab:taufinderopt}. The selection criteria are applied after the overlay mitigation and are as follows:
\begin{itemize}
\item N Leptons $\geq 1$
\item $E_{\text{vis}} < 500$ GeV 
\item Track Multiplicity $> 10$ 
\item  $-q\cos\theta_W > -0.95$ 
\item $40<M_{qq}<120$ GeV
\item $E_{\text{com}} > 100$ GeV
\item Visible $P_T > 5$ GeV  
\item  $(m^{\text{vis}}_{\text{recoil}})^2 < 135,000 \, \, \text{GeV}^2$
\end{itemize}

\begin{figure}[htpb]
\centering
    \begin{minipage}{0.49\textwidth}
        \centering
   		\begin{tikzpicture}
        \node(a){\includegraphics[width=0.99\textwidth]{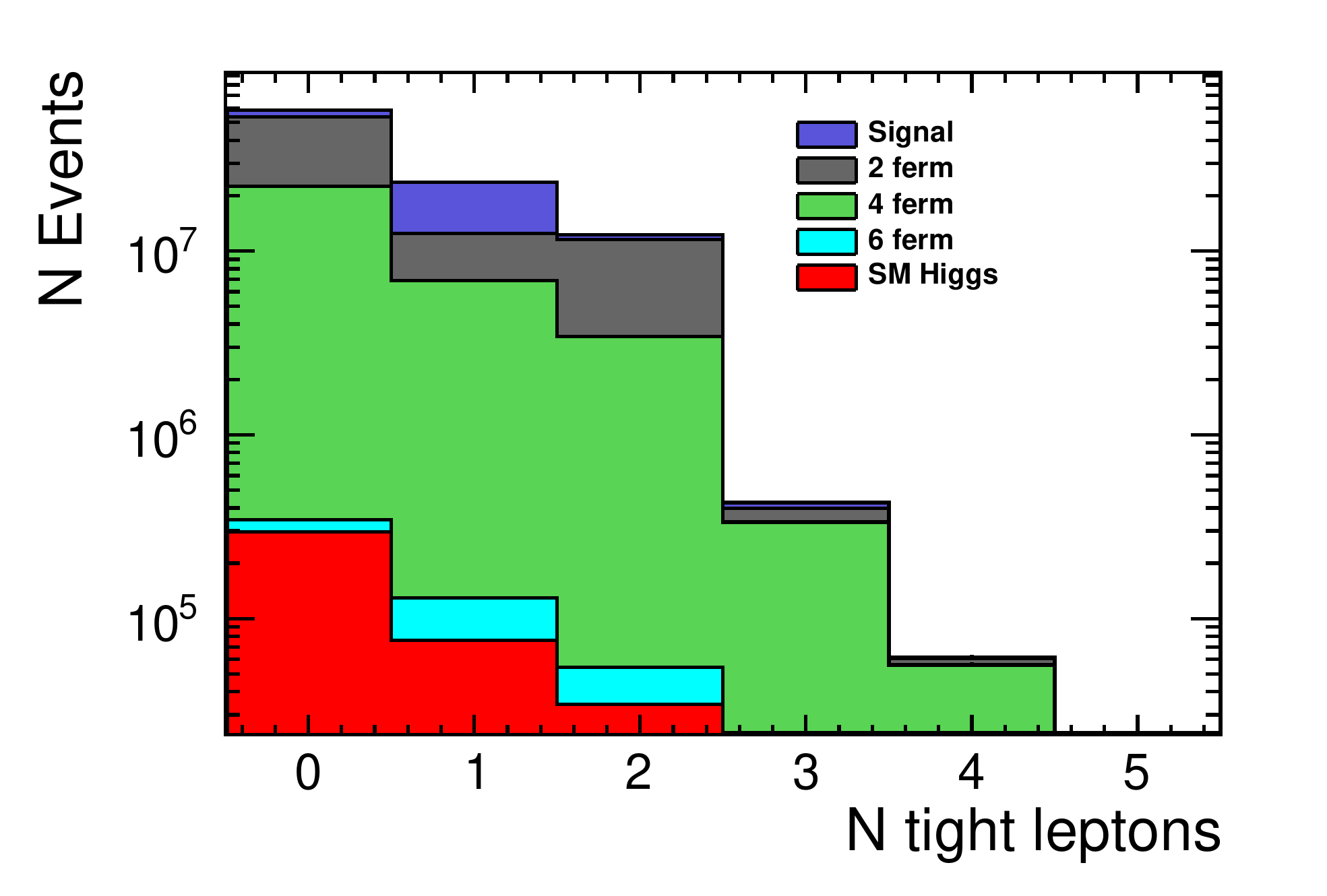}}; 
        \node at (a.north west)
        [
        anchor = center,
        xshift=6.4cm,
        yshift=-1cm
     	]
     	{
     		\includegraphics[width=0.08\textwidth]{ildlogo.png}
     	};
     	\end{tikzpicture}
    \end{minipage}\hfill
    \begin{minipage}{0.49\textwidth}
        \begin{tikzpicture}
        \node(a){\includegraphics[width=0.99\textwidth]{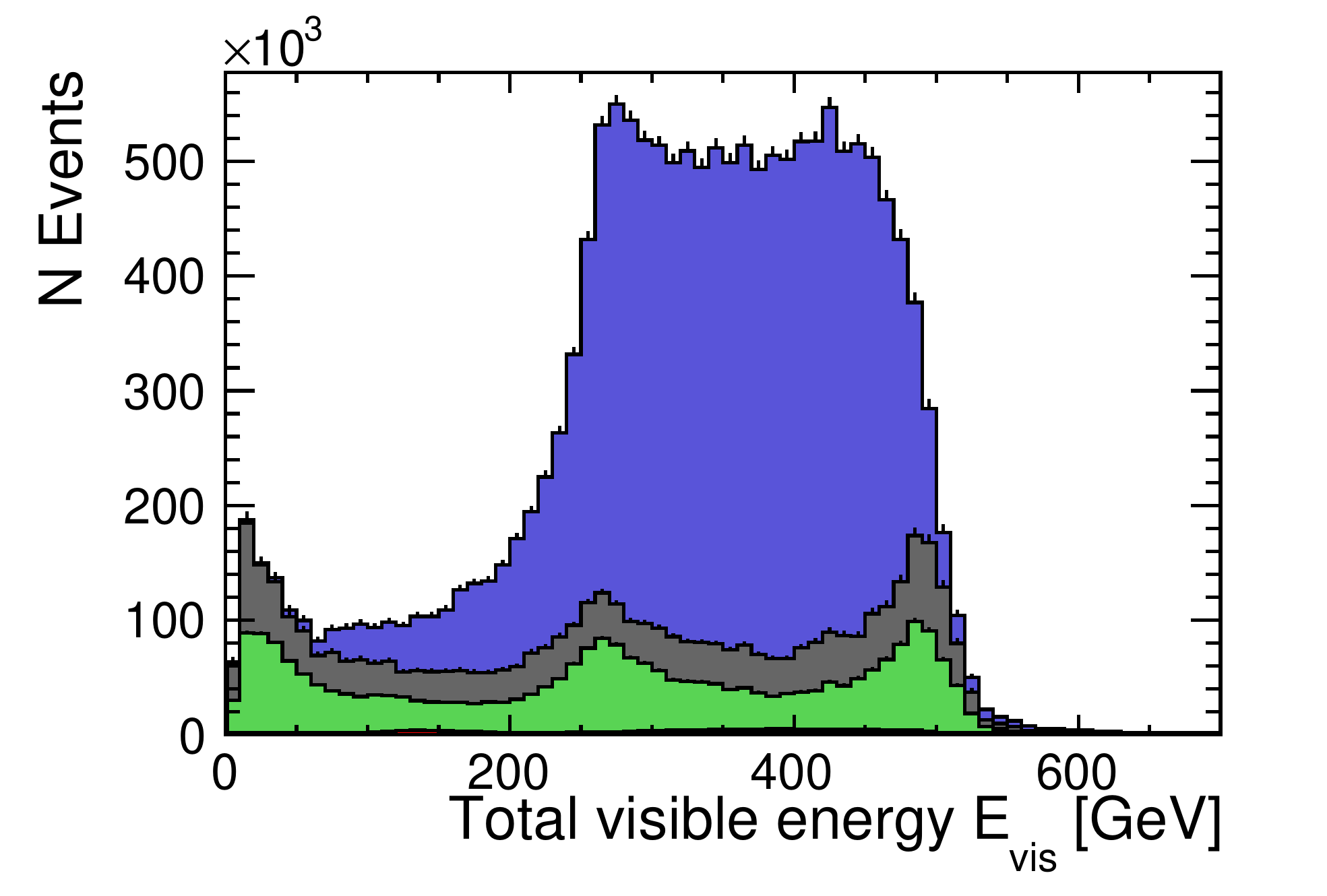}}; 
        \node at (a.north west)
        [
        anchor = center,
        xshift=6.4cm,
        yshift=-1cm
     	]
     	{
     		\includegraphics[width=0.08\textwidth]{ildlogo.png}
     	};
     	\end{tikzpicture}
     \end{minipage}\\
     \begin{minipage}{0.49\textwidth}
        \centering
   		\begin{tikzpicture}
        \node(a){\includegraphics[width=0.99\textwidth]{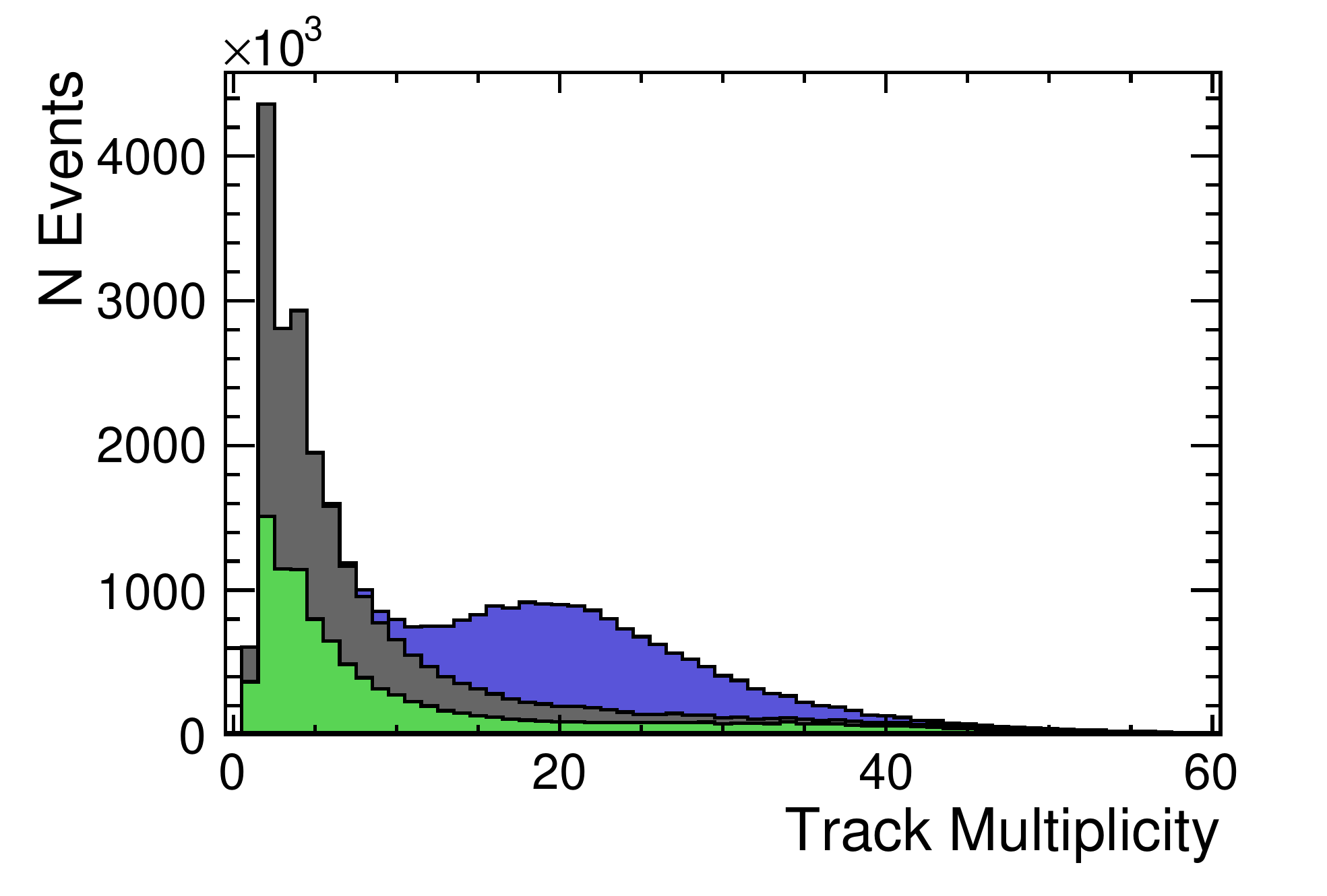}}; 
        \node at (a.north west)
        [
        anchor = center,
        xshift=6.4cm,
        yshift=-1cm
     	]
     	{
     		\includegraphics[width=0.08\textwidth]{ildlogo.png}
     	};
     	\end{tikzpicture}
    \end{minipage}\hfill
    \begin{minipage}{0.49\textwidth}
        \begin{tikzpicture}
        \node(a){\includegraphics[width=0.99\textwidth]{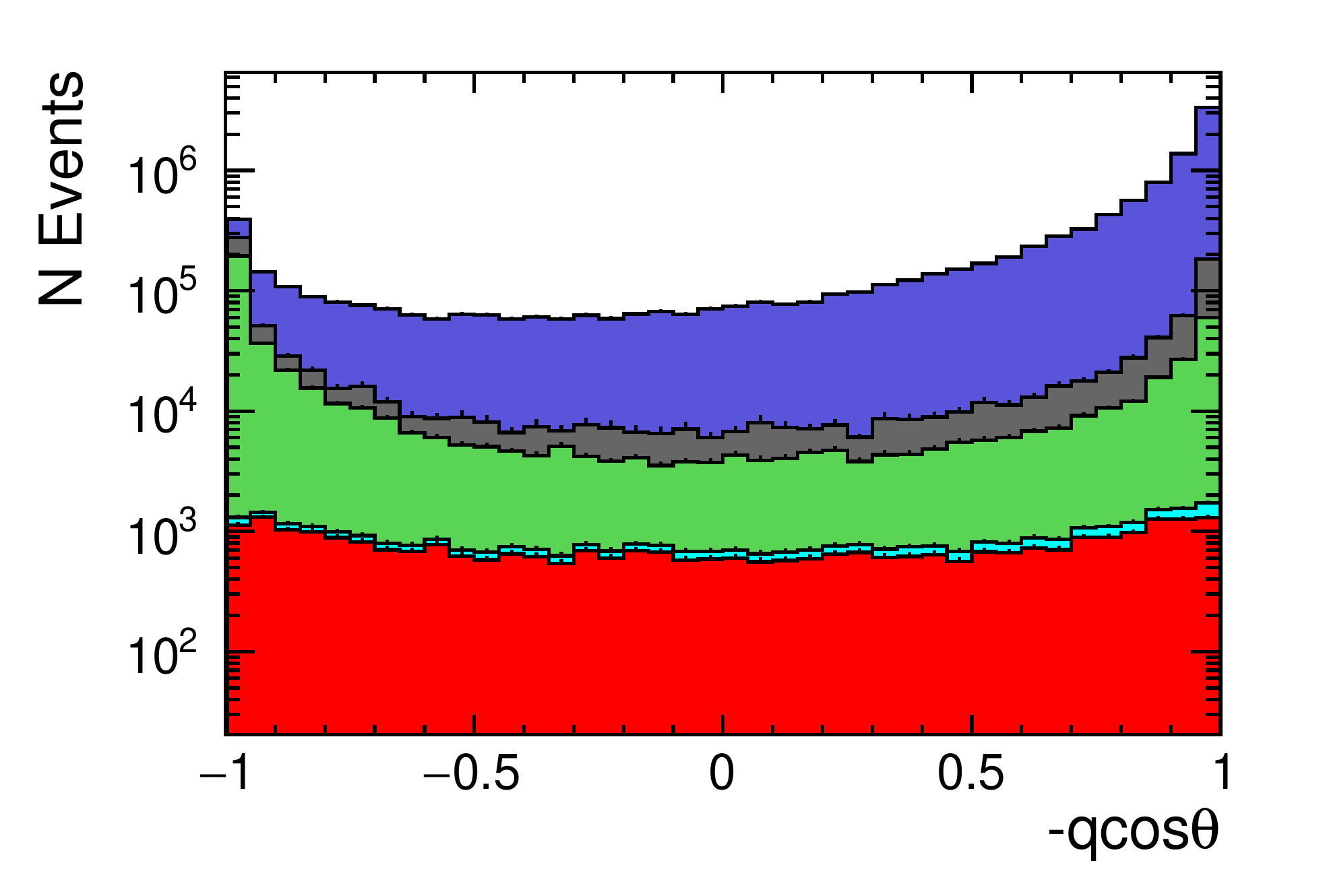}}; 
        \node at (a.north west)
        [
        anchor = center,
        xshift=6.2cm,
        yshift=-1cm
     	]
     	{
     		\includegraphics[width=0.08\textwidth]{ildlogo.png}
     	};
     	\end{tikzpicture}
     \end{minipage}\\
     	
     	\caption{ The distributions of the first four selection criteria selection criteria for the $qq\ell\nu$ tight signal region. All distributions, excluding N tight leptons, include only one preceding requirement such that there is at least 1 identified tight lepton. The signal selection only includes events that pass the tight muon cone requirement. Uses 1600 $\text{fb}^{-1}$ in $(-0.8,+0.3)$.}
     	\label{fig:cutflow}

     \end{figure}

\begin{figure}[htpb]
     \begin{minipage}{0.49\textwidth}
        \centering
   		\begin{tikzpicture}
        \node(a){\includegraphics[width=0.99\textwidth]{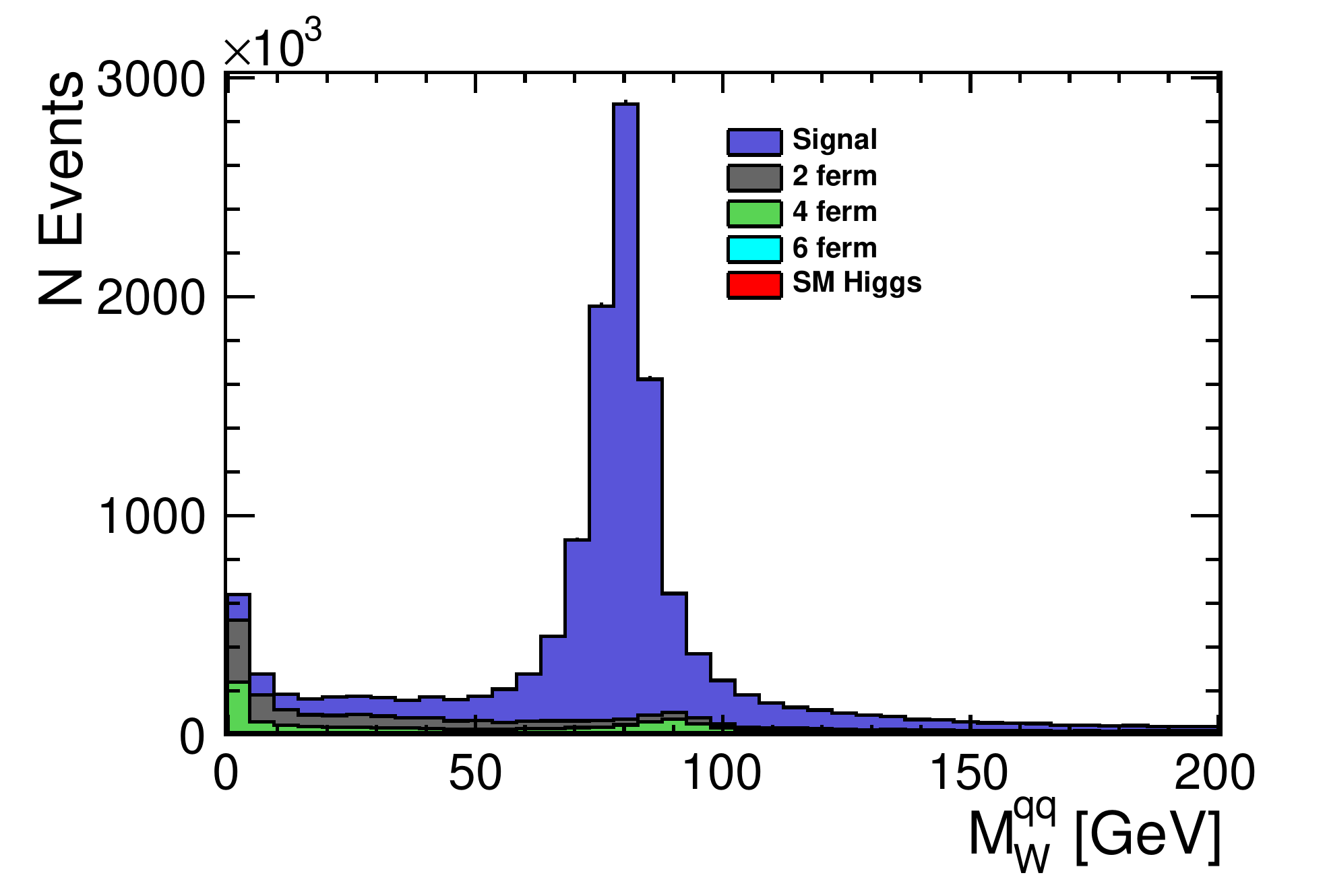}}; 
        \node at (a.north west)
        [
        anchor = center,
        xshift=6.4cm,
        yshift=-1cm
     	]
     	{
     		\includegraphics[width=0.08\textwidth]{ildlogo.png}
     	};
     	\end{tikzpicture}
    \end{minipage}\hfill
    \begin{minipage}{0.49\textwidth}
        \begin{tikzpicture}
        \node(a){\includegraphics[width=0.99\textwidth]{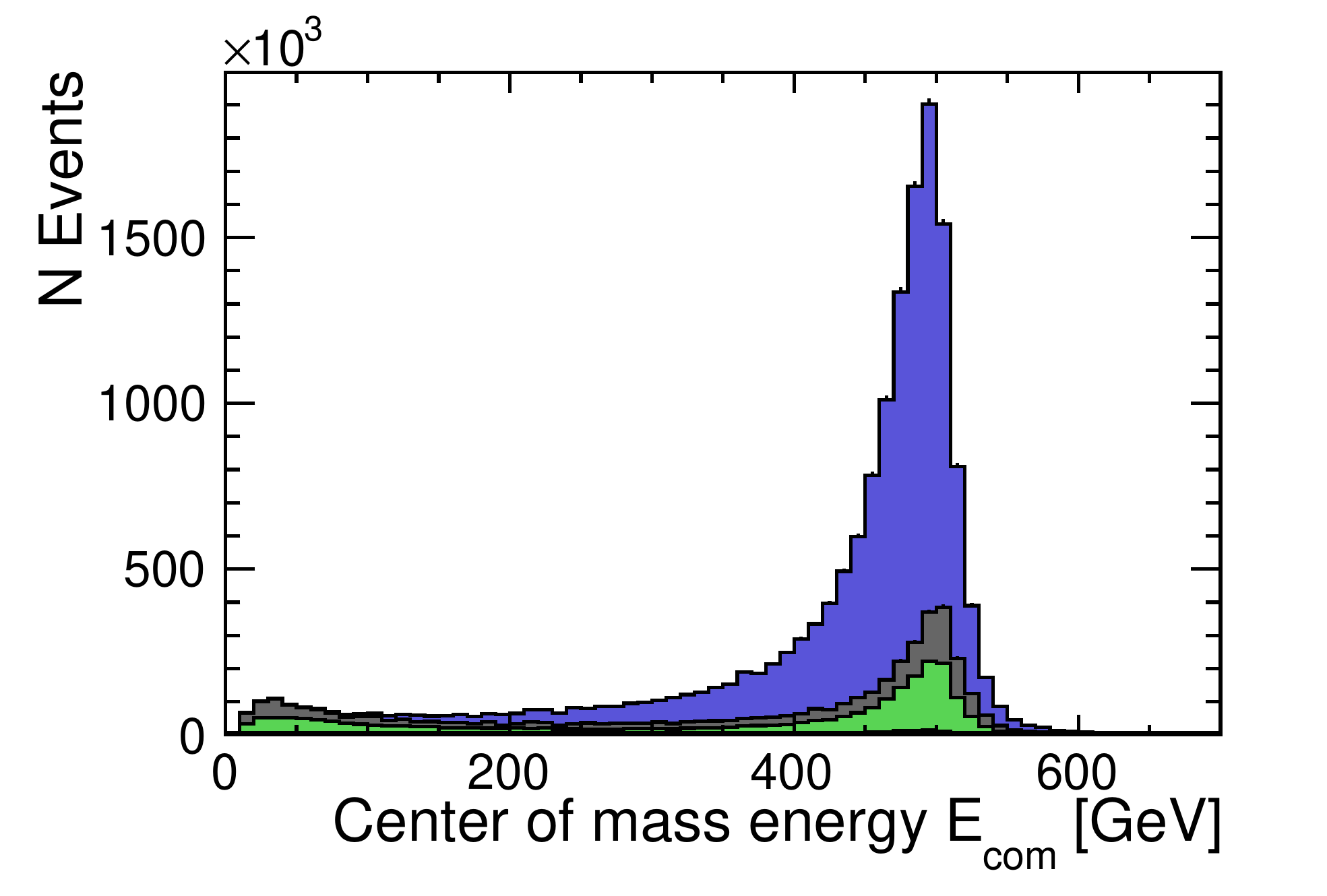}}; 
        \node at (a.north west)
        [
        anchor = center,
        xshift=6.4cm,
        yshift=-1cm
     	]
     	{
     		\includegraphics[width=0.08\textwidth]{ildlogo.png}
     	};
     	\end{tikzpicture}
     \end{minipage}\\
     \begin{minipage}{0.49\textwidth}
        \centering
   		\begin{tikzpicture}
        \node(a){\includegraphics[width=0.99\textwidth]{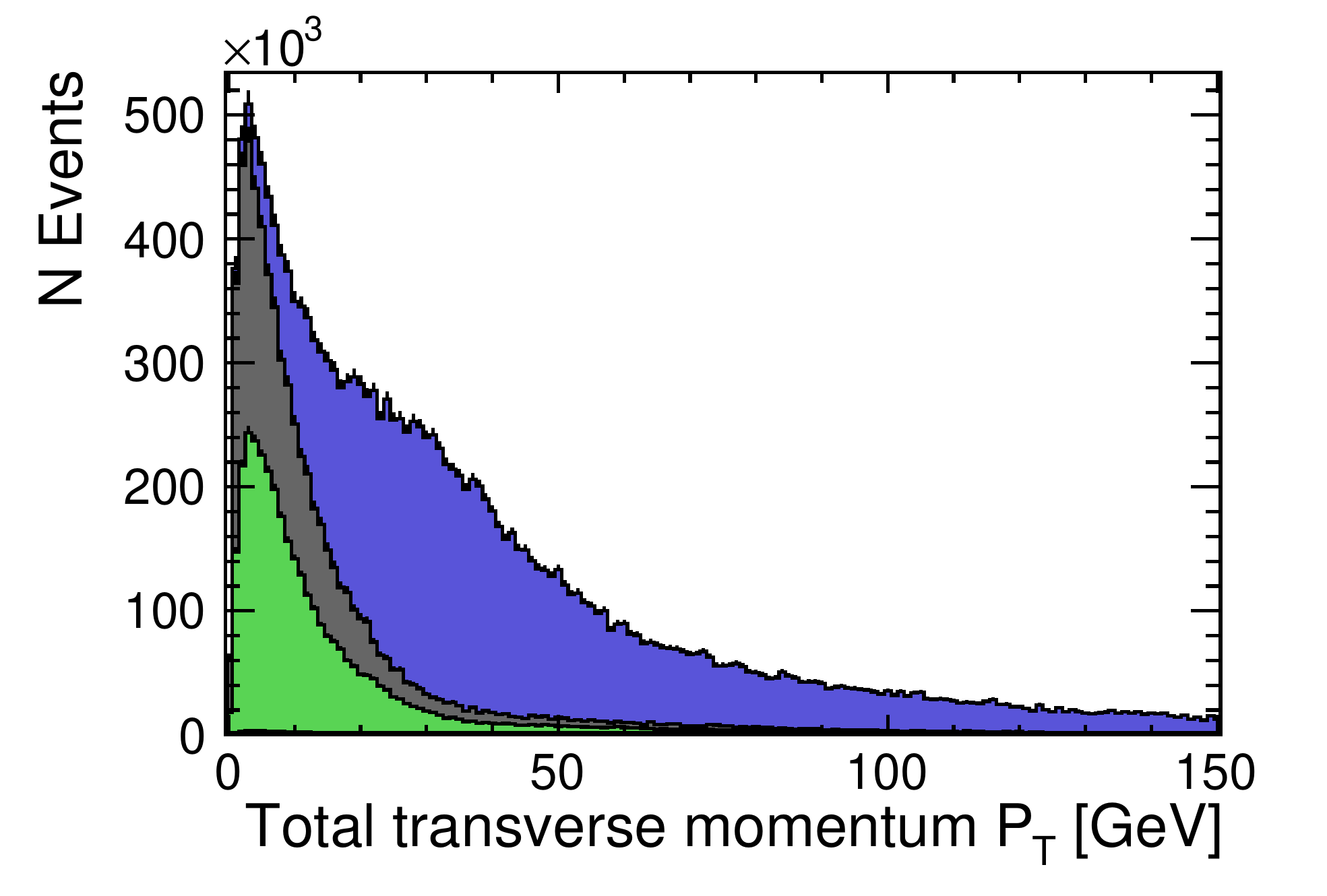}}; 
        \node at (a.north west)
        [
        anchor = center,
        xshift=6.4cm,
        yshift=-1cm
     	]
     	{
     		\includegraphics[width=0.08\textwidth]{ildlogo.png}
     	};
     	\end{tikzpicture}
    \end{minipage}\hfill
    \begin{minipage}{0.49\textwidth}
        \begin{tikzpicture}
        \node(a){\includegraphics[width=0.99\textwidth]{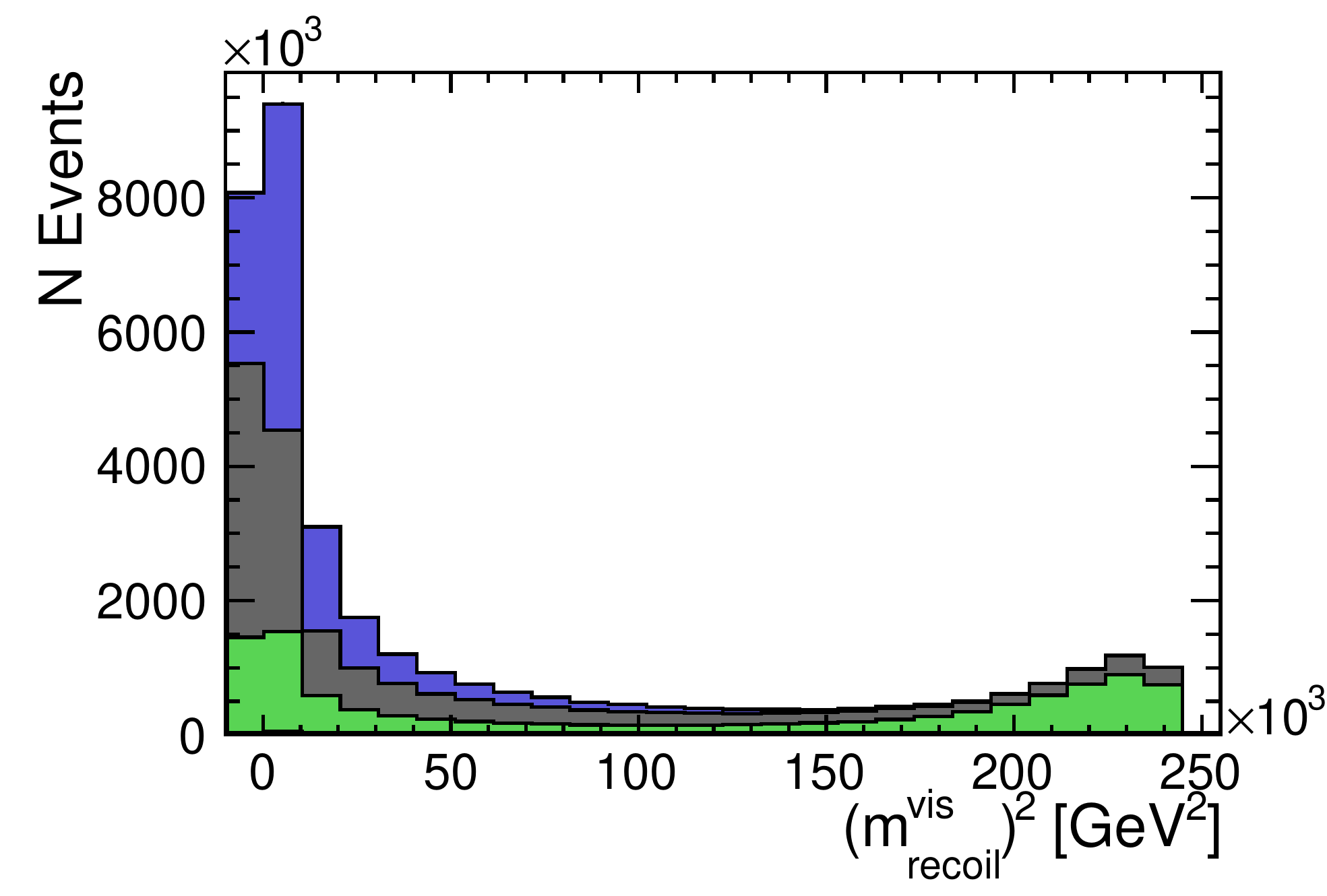}}; 
        \node at (a.north west)
        [
        anchor = center,
        xshift=6.4cm,
        yshift=-1cm
     	]
     	{
     		\includegraphics[width=0.08\textwidth]{ildlogo.png}
     	};
     	\end{tikzpicture}
     \end{minipage}\\
     	\caption{ The distributions of the last four selection criteria for the $qq\ell\nu$ tight signal region. All distributions, excluding N tight leptons, include only one preceding requirement such that there is at least 1 identified tight lepton. The signal selection only includes events that pass the tight muon cone requirement. Uses 1600 $\text{fb}^{-1}$ in $(-0.8,+0.3)$.}
     	     \label{fig:cutflow2}

\end{figure}

\begin{figure}

\centering
   \begin{minipage}{0.50\textwidth}
       \centering
      \begin{tikzpicture}
  \node(a){\includegraphics[width=0.99\textwidth]{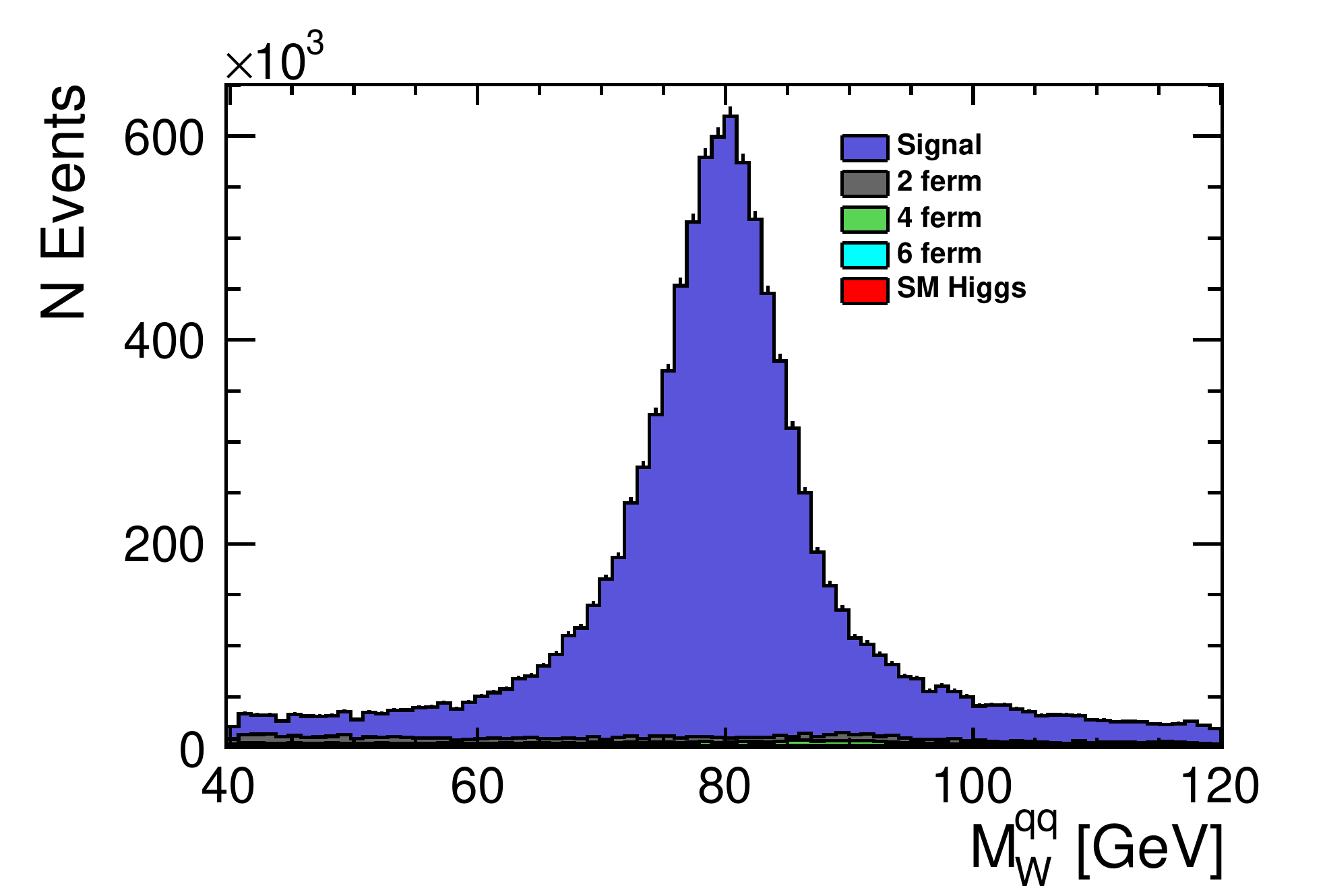}};
    \node at (a.north west)
    [
    anchor=center,
     xshift= 2.5cm,
    yshift= -1.25cm
    ]
    {
        \includegraphics[width=0.13\textwidth]{ildlogo.png}
    };
    \end{tikzpicture}

    \end{minipage}\hfill
    \begin{minipage}{0.50\textwidth}
        \centering
        \begin{tikzpicture}
      \node(a){\includegraphics[width=0.99\textwidth]{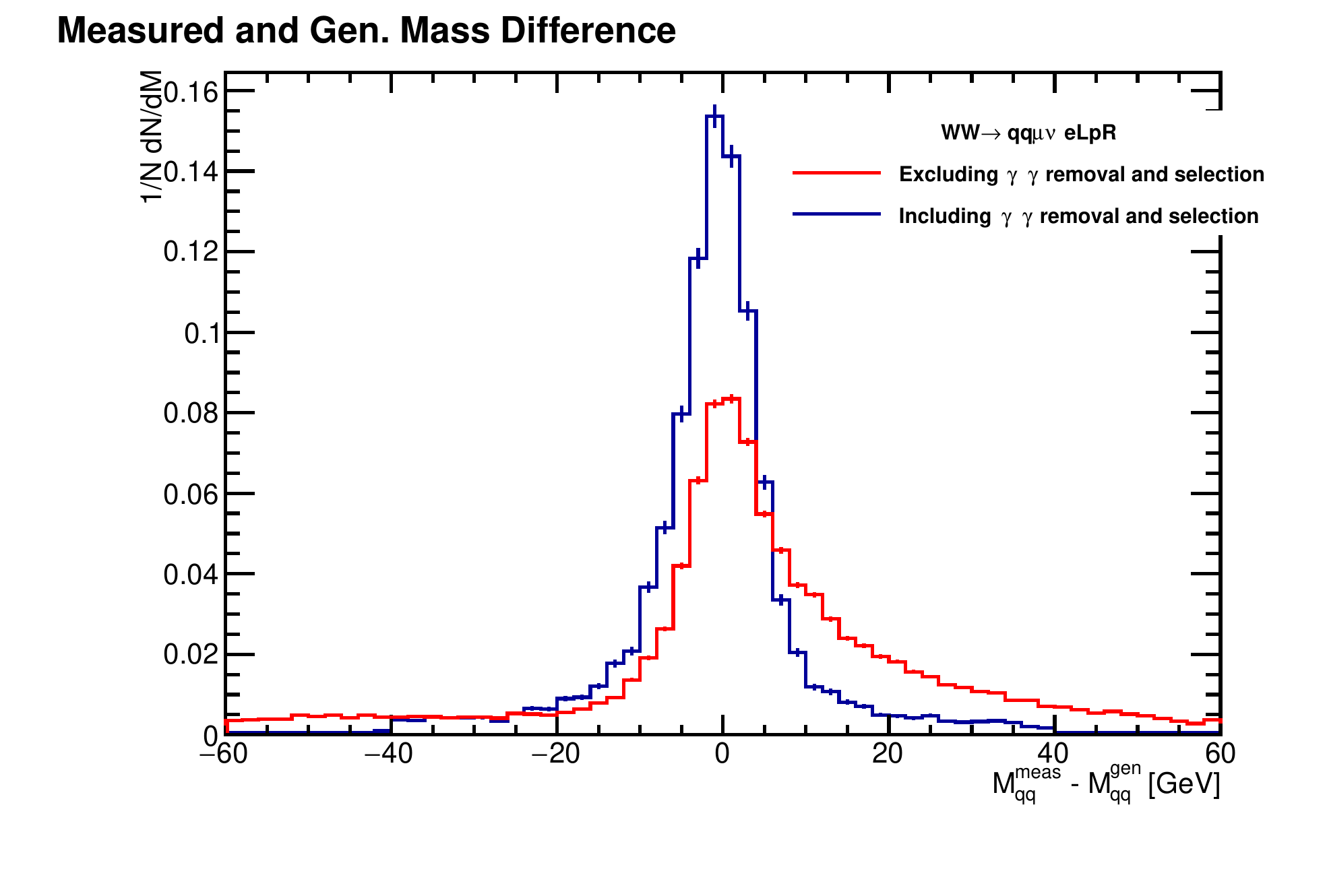}};
        \node at (a.north west)
        [
        anchor = center,
         xshift= 2.5cm,
    yshift= -1.25cm
     	]
     	{
     		\includegraphics[width=0.13\textwidth]{ildlogo.png}
     	};
    	\end{tikzpicture}
    \end{minipage}\\
     \caption{The left distribution shows the  mass of the hadronic W-boson after overlay removal and selection cuts against the remaining background events. The selection used is the tight selection. Off-shell events are not displayed. The right distribution shows the resolution of the hadronic W mass with respect to the true mass in Monte Carlo to illustrate the difference of the raw reconstruction before $\gamma \gamma$ removal and selection versus selected events proceeding $\gamma \gamma$ removal.  Uses 1600 $\text{fb}^{-1}$ in $(-0.8,+0.3)$. 
}
\label{fig:money}
\end{figure}

The track multiplicity and $E_{\text{com}}$ target 2 fermion backgrounds. $E_{\text{com}}$ is the rest-frame energy that consists of the visible and inferred missing energy. The missing energy is treated as a single neutrino with zero mass such that $E_{\text{com}} = E_{\text{vis}} + |P_{\text{miss}}| \, \,  \, \text{and} \, \, P^\mu_{\text{miss}} = (|P_{\text{miss}}| , -\sum{\vec{p}_{\text{vis}}})$. The visible $P_T$ is the magnitude of the vector of all measured transverse momenta and $E_{\text{vis}}$ is the sum of all reconstructed visible energy in an event. The $P_T$ and $E_{\text{vis}}$ cuts target processes that do not have a genuine missing energy from a neutrino. The hadronic W-mass $M_{qq}$ requirement forces the hadronic system to be ``W-like" and the recoil mass, $m_{\text{recoil}}^{\text{vis}}$, uniquely requires the visible system to be recoiling against an invisible system with little to no mass. The recoil mass is defined as $(m_{ \text{recoil}}^{\text{vis}})^2 = s + M^2_{\text{vis}} - 2\sqrt{s}E_{\text{vis}} \, \, \text{and} \, \, M^2_{\text{vis}} = ( P^{\mu}_{qq} +  P^{\mu}_{\ell})^2$. The W-scattering angle $-q\cos\theta_W$ is the angle of deflection of the system identified as $W^-$ with respect to the $e^-$ beam axis and is implemented to limit backward scattering.  The charge of the lepton is extracted from the leading momentum track from candidates with 1, 2, or 4 tracks, or in the case of 3 tracks, the charge is the sum of the three track charges. The hadronic system is then tagged with the opposite charge.  This procedure results in the correct charge assignment of the lepton before selection cuts for about $98.9\%$ of prompt muons, $94.8\%$ of prompt electrons, and $95.9\%$ of taus. Following the event selection the correct charge assignment increases to $99.9\%$ for prompt muons, $98.3\%$ for prompt electrons, and $98.8\%$ for taus. The distributions for the tight selection are shown in Figure \ref{fig:cutflow} and Figure \ref{fig:cutflow2}. The details of the tight and loose selection for $(-0.8,+0.3)$ at 1600 $\text{fb}^{-1}$ are summarized in Table \ref{tab:selection}. The selections in Table \ref{tab:selection} differ by the veto of the tight lepton in the loose selection, where the preference for any event is to always choose a tight lepton over loose. The primary selection also includes only ``WW-like" signal, this type of signal is such that both of the true fermion pairs invariant masses are each within $\pm10$ GeV of the nominal W mass. If an event contains a fermion pair that is outside the WW-like range, it is designated as an off-shell (O.S.) event and is placed in a different category. The selection details for events from the O.S. category are given in Table \ref{tab:os}. The selection criterion applied to every category are optimized to maximize the efficiency and purity of the total signal for the tight selection with WW-like events. The results are summarized with the two selection cones with and without O.S. events in Table \ref{tab:summary}. The final results for the selection with no restrictions on W mass and both cones, yields a signal efficiency of $60\%$ with only $7\%$ of background contaminating the selected events. The W-boson invariant mass distribution after selection is shown with mass resolution in Figure \ref{fig:money}.

\begin{table}

\caption{The tight and loose selection for 1600 $\text{fb}^{-1}$ $(-0.8,+0.3)$ WW-like events. The tight selection is most efficient with prompt leptons with $69\%$ and $65\%$  for muons and electrons respectively, but struggles to efficiently reconstruct taus. The loose selection recovers $10\%$ of the tau efficiency and is inefficient for prompt leptons because of the tight lepton veto. The tight lepton veto enforces the orthogonality of the selections and gives preference toward better lepton candidates. The signal categories Base Evts. only include events in which the true W mass of both fermion pairs are within 10 GeV of the nominal W mass.}
\label{tab:selection}
 \scriptsize
 Tight selection with muon cone \\
   \begin{tabular}{|p{0.11\textwidth}|p{0.06\textwidth}p{0.06\textwidth}p{0.06\textwidth}|p{0.08\textwidth}|p{0.06\textwidth}p{0.06\textwidth}p{0.06\textwidth}p{0.06\textwidth}|p{0.08\textwidth}|}
\hline 
 $(\times 10^5)$  & $qq\mu\nu$ & $qqe\nu$ & $qq\tau\nu$ & Tot. Sig. & 2f & 4f & 6f & Higgs & Tot. Bkg. \\ \hline 
Base Evts & {38.7 } &  {38.9 } &  {39.0} & {117} &  {422} &  {322} &  {2.14} &  {4.12} & 750 \\ 

Lep. + Cone & {33.1 } &  {32.0 } &  {22.8} & {87.8} &  {115} &  {118} &  {1.63} &  {1.15} & 236 \\ 
 
$E_{\text{vis}}$ & {32.8 } &  {31.1 } &  {22.7} & {86.7} &  {106} &  {115} &  {1.62} &  {1.11} & 224\\ 
 
N Tracks & {31.9 } &  {30.3 } &  {22.1} & {84.3} &  {25.4} &  {25.9} &  {1.49} &  {0.889} & 53.7\\ 
 
$-q\cos\theta$ & {31.8 } &  {30.1 } &  {21.8} & {83.7} &  {21.9} &  {22.6} &  {1.44} &  {0.852} & 46.8\\ 
 
$M_{qq}$ $>$40 & {29.4 } &  {28.0 } &  {20.3} & {77.7} &  {11.3} &  {13.3} &  {1.42} &  {0.756} & 26.8\\ 
 
$M_{qq}$ $<$120 & {27.2 } &  {25.7 } &  {18.3} & {71.3} &  {5.68} &  {2.68} &  {0.202} &  {0.297} & 8.86\\ 
 
$E_{\text{com}}$ & {27.2 } &  {25.7} &  {18.3} & {71.3} &  {5.58} &  {2.65} &  {0.202} &  {0.296} & 8.73\\ 

$P_T$ vis. & {26.9 } &  {25.5} &  {18.1} & {70.5} &  {3.21} &  {2.37} &  {0.201} &  {0.294} & 6.08\\ 

$(m^{\text{vis}}_{\text{recoil}})^2$ & {26.9 } &  {25.4 } &  {18.0} & {70.3} &  {2.93} &  {2.02} &  {0.194} &  {0.223} & 5.36 \\ 
\hline 

 $\epsilon \, \, (\%)$ & $69.4 $ & $65.4 $ & $46.2$ &  $60.3 $ & $0.69 $ & $0.626 $ & $9.05 $ & $5.41 $ & $0.72$ \\ 
 
 			& $\pm 0.2$ & $\pm 0.2$ & $\pm 0.3$ & $\pm 0.2$ & $\pm 0.01$ & $\pm 0.008 $& $\pm 0.02$ & $\pm0.05$  & $\pm 0.01$\\
\hline
\end{tabular}
\quad \quad \\
Loose selection with tau cone\\
\begin{tabular}{|p{0.11\textwidth}|p{0.06\textwidth}p{0.06\textwidth}p{0.06\textwidth}|p{0.08\textwidth}|p{0.06\textwidth}p{0.06\textwidth}p{0.06\textwidth}p{0.06\textwidth}|p{0.08\textwidth}|}
\hline 
 $(\times 10^5)$  & $qq\mu\nu$ & $qqe\nu$ & $qq\tau\nu$ & Tot. Sig. & 2f & 4f & 6f & Higgs & Tot. Bkg. \\ \hline 
Base Evts & {38.7 } &  {38.9 } &  {39.0} & {117} &  {422} &  {322} &  {2.14} &  {4.12} & 750\\ 
 
Lep.+ Cone & {33.6 } &  {33.0 } &  {28.2} & {94.8} &  {130} &  {136} &  {1.77} &  {1.38} & 269\\ 

Tight Veto & {0.772 } &  {1.28 } &  {5.70} & {7.76} &  {19.3} &  {21.5} &  {0.161} &  {0.312} & 41.3 \\ 
 
$E_{\text{vis}}$ & {0.764 } &  {1.26 } &  {5.70} & {7.72} &  {18.2} &  {19.4} &  {0.154} &  {0.302} & 38.1\\ 

N Tracks & {0.737 } &  {1.21 } &  {5.54} & {7.49} &  {15.0} &  {16.4} &  {0.151} &  {0.271} & 31.8\\ 
 
$-q\cos\theta$ & {0.630 } &  {1.12 } &  {5.32} & {7.07} &  {11.1} &  {14.1} &  {0.145} &  {0.256} &25.6\\ 
 
$M_{qq}$ $>$ 40 & {0.492 } &  {0.972 } &  {4.86} & {6.33} &  {5.98} &  {13.0} &  {0.144} &  {0.233}& 19.4 \\ 

$M_{qq}$ $<$ 120 & {0.404 } &  {0.781 } &  {4.16} & {5.35} &  {2.58} &  {1.11} &  {0.0111} &  {0.124} & 3.83\\ 
 
$E_{\text{com}}$ & {0.404 } &  {0.781 } &  {4.16} & {5.34} &  {2.50} &  {1.10} &  {0.0111} &  {0.124} &3.74\\ 

$P_T$ vis. & {0.400 } &  {0.774 } &  {4.12} & {5.29} &  {1.17} &  {1.01} &  {0.0111} &  {0.123} &2.31\\ 
 
$(m^{\text{vis}}_{\text{recoil}})^2$ & {0.394 } &  {0.770 } &  {4.07} & {5.24} &  {1.02} &  {0.759} &  {0.0102} &  {0.0973}& 1.89 \\ 
\hline 
 $\epsilon \, \, (\%)$ & $1.02 $ & $1.98 $ & $10.5 $ &  $4.49 $ & $0.241 $ & $0.236 $ & $0.474 $ & $2.36 $ & $0.251$\\ 

  	     & $\pm 0.05$ & $\pm 0.07$ & $\pm 0.2$ & $\pm 0.07$ & $\pm 0.003$ & $\pm 0.004$ & $\pm 0.007$ & $\pm 0.02$ & $\pm 0.003$ \\

 \hline
 \end{tabular}

\end{table}

\begin{table}
\centering
\caption{The tight and loose selection for 1600 $\text{fb}^{-1} (-0.8,+0.3)$ non WW-like or off-shell (O.S.) events.  The Base
Evts. only include events in which the true W mass of both fermion pairs has a minimum absolute difference
with the nominal W mass of 10 GeV. The selection is not optimized for selecting these types of events so the overall efficiency is less, compared to the WW-like events.}
\begin{minipage}{0.49\linewidth}
\scriptsize
 Tight selection with muon cone \\
\begin{tabular}{|p{0.24\textwidth}|p{0.14\textwidth}p{0.14\textwidth}p{0.14\textwidth}|p{0.14\textwidth}|}
\hline
  $(\times 10^5)$& $qq\mu\nu$ & $qqe\nu$ & $qq\tau\nu$ & Tot. Sig. \\ 
\hline
Base Evts &{5.78 } & {38.8 } & {5.70} & 50.3\\ 
 
Lep. + Cone &{5.11 } & {22.7 } & {3.42} & 31.2\\ 

$E_{\text{vis}}$ &{5.08 } & {22.5 } & {3.41} & 31.0\\ 

N Tracks &{4.95 } & {21.9 } & {3.35}& 30.2\\ 
 
$-q\cos\theta$ &{4.94 } & {21.0 } & {3.31}& 29.3\\ 
 
$M_{qq}>40$ &{4.67} & {20.1 } & {3.14} & 27.9\\ 
 
$M_{qq}<120$ &{3.44 } & {18.1 } & {2.39} & 23.9\\ 

$E_{\text{com}}$ &{3.44 } & {18.1 } & {2.39} & 23.9\\ 
 
$P_T$ vis. &{3.40 } & {18.0 } & {2.36} & 23.8\\ 
 
$(m^{\text{vis}}_{\text{recoil}})^2$ & {3.40 } & {17.6 } & {2.34}& 23.3\\ 
\hline 
 $\epsilon \, \, (\%)$ & $58.8$ & $45.4$ & $41.1$ & $46.3$ \\ 
 						& $\pm 0.6$ & $\pm 0.1$ & $\pm 0.6$ & $\pm 0.1$\\
 \hline
\end{tabular}
\end{minipage}


\begin{minipage}{0.49\linewidth}
\scriptsize
\quad \quad \\
Loose selection with tau cone\\
 \begin{tabular}{|p{0.24\textwidth}|p{0.14\textwidth}p{0.14\textwidth}p{0.13\textwidth}|p{0.14\textwidth}|}
\hline 
 $(\times 10^5) $ & $qq\mu\nu$  & $qqe\nu$ & $qq\tau\nu$ & Tot. Sig.\\ \hline 
Base Evts &{5.78 } & {38.8 } & {5.70} & 50.3\\ 
 
Lep.+ Cone &{5.15 } & {24.7 } & {4.26} & 34.1\\ 

Tight Veto &{0.082 } & {2.61 } & {0.88} & 3.57\\ 

$E_{\text{vis}}$ &{0.079 } & {2.61 } & {0.88} & 3.57\\ 

N Tracks &{0.076 } & {2.48 } & {0.86} & 3.42\\ 

$-q\cos\theta$ &{0.070 } & {2.31 } & {0.82} & 3.20\\ 
 
$M_{qq} > 40$ &{0.056 } & {1.38 } & {0.77} & 2.20\\ 

$M_{qq} < 120$ &{0.036 } & {1.20 } & {0.48} & 2.05\\ 
 
$E_{\text{com}}$ &{0.036 } & {1.18 } & {0.48} & 2.03\\ 
 
$P_T$ vis. &{0.036 } & {1.18 } & {0.48 }& 2.02\\ 

$(m^{\text{vis}}_{\text{recoil}})^2$ &{0.036 } & {0.92 } & {0.47} & 1.42\\ 
\hline 
 $\epsilon \, \, (\%)$ & $0.63 $ & $2.36$ & $8.3$ & $2.82$\\ 
  			& $\pm 0.1$ & $\pm 0.05$ & $\pm 0.4$ & $\pm 0.05$\\  
 \hline
 \end{tabular}

\end{minipage}
\label{tab:os}
 \end{table}

\begin{table}
\caption{Selection summary showing the number of background events that pass the event selection, efficiency, and purity for the tight selection or combined selections and with or without the off-shell contributions.  The O.S. contributions are less efficiently reconstructed so their inclusion reduces the overall signal efficiency but boosts the purity by increasing the base number of signal events that can be selected. Selection is performed with 1600 $\text{fb}^{-1}$ in $(-0.8, +0.3)$. }
\label{tab:summary}
 \begin{tabular}{ |p{0.12\textwidth}|p{0.12\textwidth}p{0.15\textwidth}|p{0.08\textwidth}|p{0.12\textwidth}p{0.17\textwidth}|p{0.08\textwidth}|} 
 \hline 
   &  \multicolumn{3}{|l|}{Tight Selection} &  \multicolumn{3}{|l|}{ Tight + Loose Sel.}  \\  \hline  
 & Sel. Total $(\times 10^5)$ & Efficiency $(\%)$ & Purity $(\%)$ & Sel. Total $(\times 10^5)$ & Efficiency $\,\,\,\,$ $(\%)$& Purity $(\%)$ \\ 
 \hline  
 Bkg. &  {5.36} & 0.72 $\pm$ 0.01 & --&  {7.25} & 0.967 $\pm$ 0.005 &--  \\ 
 Signal &  {70.3} & 60.3 $\pm$ 0.2 & 92.9  &  {75.5} & 64.6 $\pm$ 0.2 & 91.2 \\ 
 Sig.+O.S. &  {93.6} & 56.0 $\pm$ 0.1 & 94.6 &  {100.3} & 60.0 $\pm$ 0.1 & 93.3 \\ 
\hline 
\end{tabular} 
\end{table}


\subsection{Results}
\label{subsec:wmass}

\begin{figure}

\centering
    \begin{minipage}{0.49\textwidth}
        \centering
   		\begin{tikzpicture}
        \node(a){\includegraphics[width=0.99\textwidth]{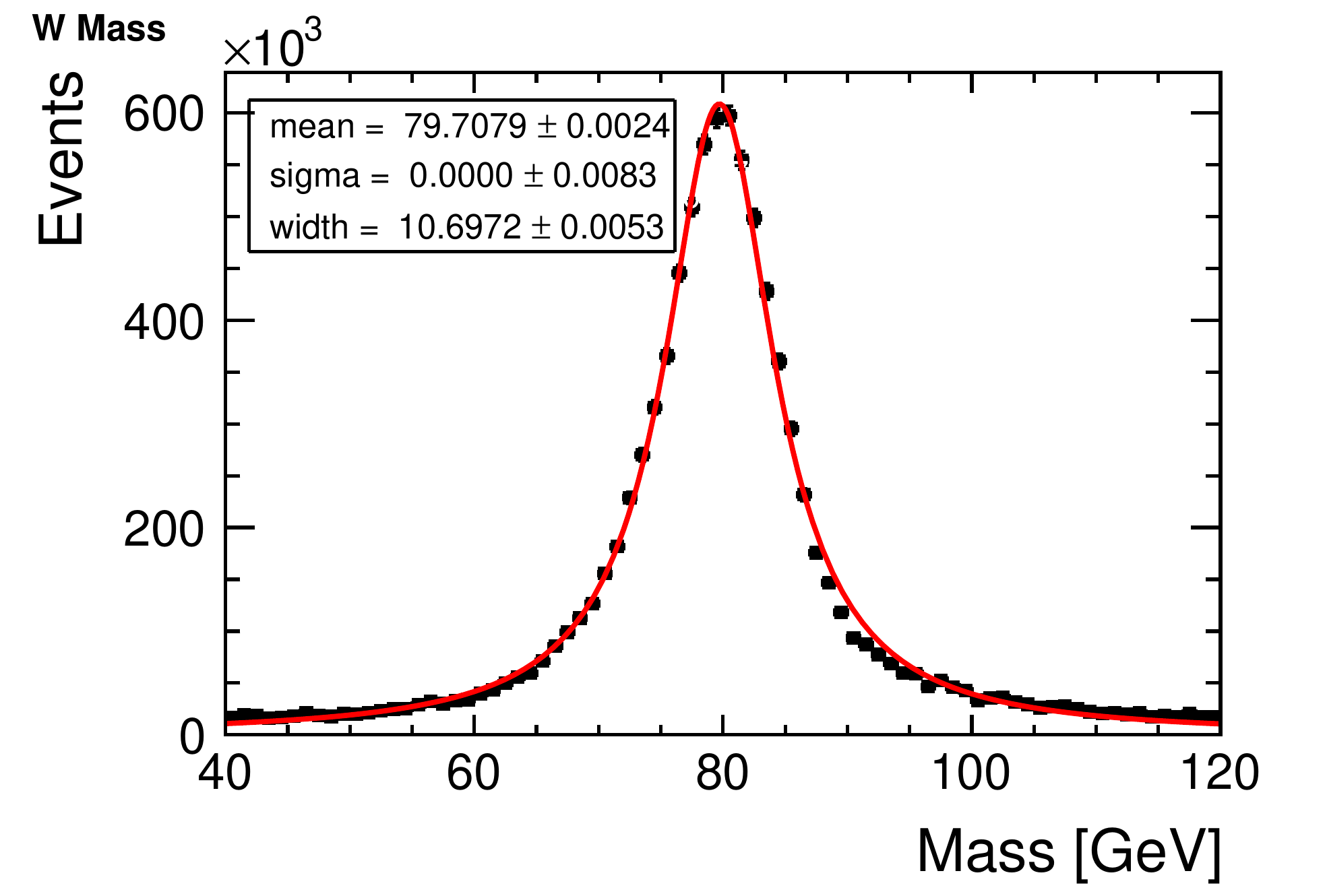}}; 
        \node at (a.north west)
        [
        anchor = center,
        xshift=6cm,
        yshift=-1.25cm
     	]
     	{
     		\includegraphics[width=0.13\textwidth]{ildlogo.png}
     	};
     	\end{tikzpicture}
    \end{minipage}\hfill
    \begin{minipage}{0.49\textwidth}
        \centering
        \includegraphics[width=0.99\textwidth]{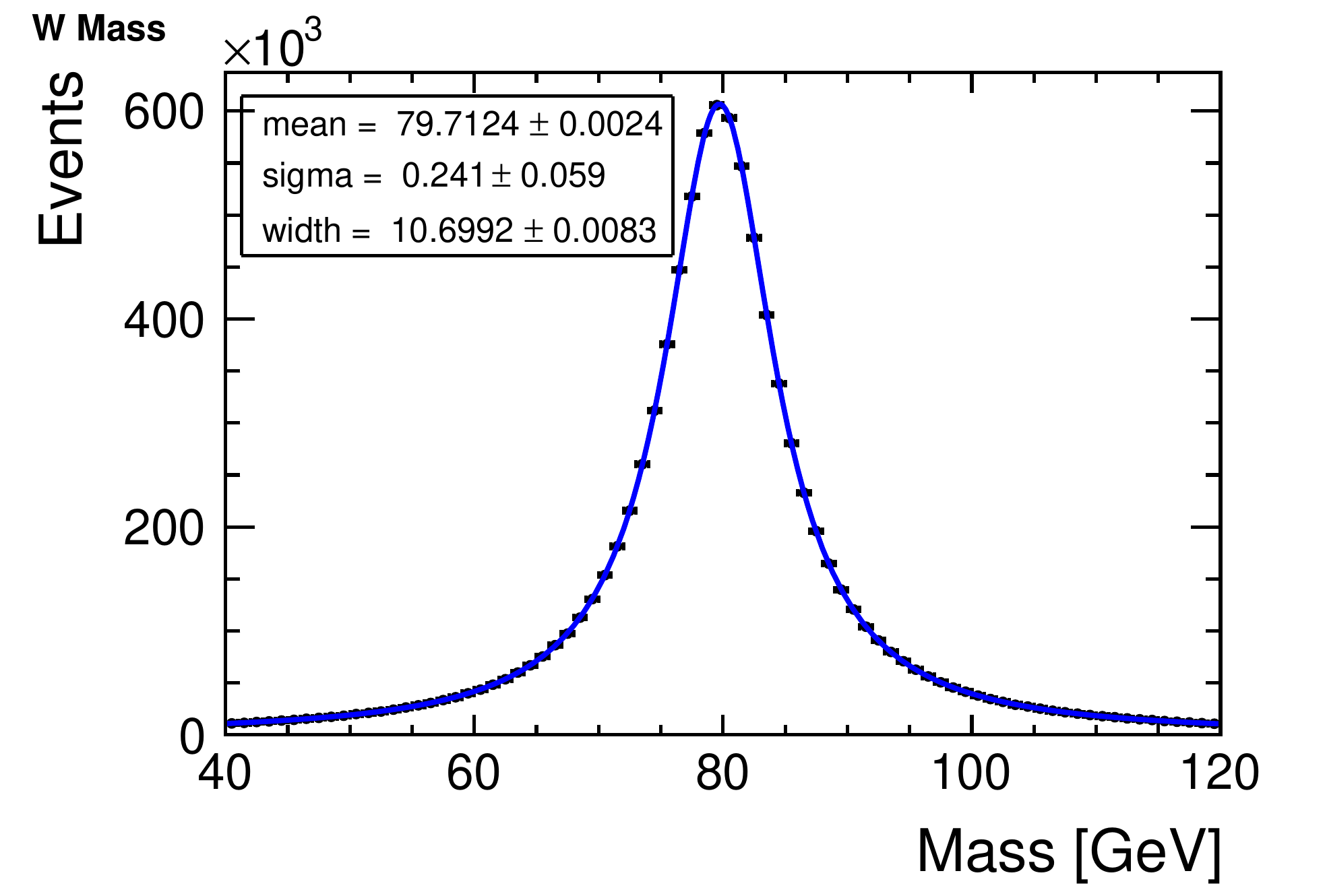} 
     
     \end{minipage}\\
     \caption{ The left distribution shows the $(-0.8,+0.3)$ W mass distribution for all signal after selection cuts. The fit results in the Voigtian parameters $M_W = 79.7079$ GeV, width $\Gamma_W = 10.6972$ GeV and $\sigma_W = 0.0$ GeV with Monte Carlo statistics scaled up to 1600 $\text{fb}^{-1}$. The right distribution shows the refit with 9.36M W's generated according to the fitted model.}
\label{fig:badfit}

\end{figure}

A fit with the convolution of a Breit-Wigner with a Gaussian (Voigtian) of the hadronic W mass is performed on the tight signal sample with the combined lepton categories shown in Figure \ref{fig:badfit}. The resulting fit models the shape and the mean of the distribution well but deviates around 90 GeV and the edges of the fit window. The width of the fit is also in excess of the true width, which is about 2 GeV. This means that the Voigtian model is inadequate in describing the simulated data likely due to the inadequacy of a single Gaussian for modeling the resolution. However, because the shape is similar to simulated data, the fitted model is used to understand the achievable mass resolution given a perfect model. Statistics consistent with 1600 $\text{fb}^{-1}$ (9.36M Events) are produced according to the previously fitted model with a mean  $M_W = 79.7079$ GeV, width $\Gamma_W = 10.6972$ GeV and $\sigma_W = 0.0$ GeV and refitted to achieve the statistical error on the mean $\Delta M_W \text{(stat.)} = 2.4$ MeV with goodness-of-fit $\chi^2 / ndof = 67.8/77$. This fit neglects background contributions but includes the off shell contributions. The toy model refit is also included in Figure \ref{fig:badfit}.
The cross-section and errors are extracted from Table \ref{tab:selection} according to the formula:
\begin{equation}
\sigma = \frac{N_T - N_B}{L \epsilon}
\end{equation}
where $N_T$ is the observed number of events that passes the selection and $N_B$ is the expected number of background events that contaminate the signal selection. The resulting statistical error on the cross-section is dominated by  Poisson errors on $N_T$, with sub leading contributions from L such that $\Delta L/L = 0.026 \%$ and polarization uncertainty contributing $0.0019\%$ overall to $\Delta L/L$ \cite{ilcluminosity}. With $\epsilon$, $L$, and $N_B$ known perfectly, the combination of the errors for efficiency and $N_T$, and neglecting $\Delta L$, the resulting cross-section obtained in the combined selection is $8001 \pm 17 \, \, \text{fb}$ with statistical error of $\Delta \sigma/\sigma = 0.036 \%$. 




\section{Discussion and Conclusions}
\label{conclude}
The results show a promising start to potential electroweak precision measurements for the ILC with the statistical error on the mass of $\Delta M_W\text{(stat.)}= 2.4$ MeV from a full detector simulation study. Such a measurement is very competitive with the current PDG measurement of $\Delta M_W=12$ MeV, assuming that systematic uncertainties can be sufficiently controlled.  Another ILC study of the W mass showed that the achievable statistical error is $\Delta M_W\text{(stat.)} = 1.6$ MeV assuming an effective Gaussian mass resolution of $\sigma_W= 4$ GeV convolved with the intrinsic width per $10^7$ hadronic W's\cite{graham}. The mass measurement at ILC500 is more challenging because the W pair overlapping with overlay is not a major issue at lower center of mass energies. The equivalent precision may be unobtainable at higher $\sqrt{s}$, but propagating an estimated systematic uncertainty on the W mass for this study of $4.0$ MeV yields a total uncertainty of $\Delta M_W = 4.7$ MeV. Both measurements, however, are dominated by the systematic uncertainties from the effective jet energy scale which is a challenging demand.  The statistical error on the cross-section also shows the utility of semileptonic WW as a method to precisely measure the beam polarization at the interaction point. This offers an important alternative for  correctly measuring processes that are sensitive to beam polarization and assists in quantifying beam depolarization from collisions.

The lepton identification and charge assignment performance is exceptional with an overall correct charge assignment of $98.8\%$ over all three channels. This has the biggest impact on the measurement of charged triple gauge couplings that rely on the identification of the $W^-$. The lepton identification itself, still has room for improvement in ways such as (1) a multivariate type approach with TauFinder and (2) re-optimization of the parameters over a mixed polarization beam-scenario which involves both LR and RL events.  The optimization of the TauFinder parameters also led to a choice of 150 mrad for the isolation cone for each category. However, allowing the isolation to grow wider could mean over tuning the lepton identification to the topologies specific to LR. Alternative tau finding tools could also be considered and optimized for this analysis e.g. TaJet which has been developed for $\tau\tau$ analyses \cite{tajet}.

The pileup mitigation is a mostly unexplored avenue of reconstruction in ILC500 as most processes are produced centrally. The techniques developed to remove overlay are optimized for W mass measurement, but, can be easily adapted to general usage in any type of process where the standard approaches for overlay removal are inadequate. 

The event selection can also be improved. Additional cuts were explored such as the leptonic W mass, or maximum track multiplicity. The leptonic W mass cut best motivates the categorization approach of WW-like and not WW-like types of event, but, this cut, and others mentioned do not improve the overall efficiency times purity of the analysis.  Specifically, the leptonic W mass can be improved and applied to event selection by using a more sophisticated calculation for the neutrino momentum. This can be done taking into account potential ISR in the $z$-direction, and applying a kinematic fit with constraints on the energy, momentum, and equal W masses. These adjustments would significantly improve the measured leptonic W mass and enhance the performance of the event selection, but requires well modeled uncertainties, and was beyond the scope of the present work. 
 
Some additional detector benchmarks can immediately follow the results of this analysis, one would be the quality of separation between prompt muons and secondary muons from tau decay to evaluate the performance of the vertex detector. Another study that should be done is examination of the analysis efficiency as a function of the polar angle, which tests the performance of the forward calorimeters.  Overall, the semileptonic analysis offers keen insights to analysis performed in an electron positron collider with $\sqrt{s} = 500$ GeV. The statistical errors on cross-section and mass are the first step but an important and tractable step in electroweak precision measurement at the ILC.  The semileptonic channel still offers significantly more important physics in terms of TGC and polarization measurements in addition to expanding and improving the analysis presented here. 



\Acknowledgments
Thanks for comments and advice related to this project from Graham Wilson, Klaus Desch, and Jenny List. I would also like to thank the LCC generator working group and the ILD software working group for providing the simulation and reconstruction tools and producing the Monte Carlo samples used in this study.
This work has benefited from computing services provided by the ILC Virtual Organization, supported by the national resource providers of the EGI Federation and the Open Science GRID


\bibliographystyle{utphys}
\bibliography{mybib}

\end{document}